\documentclass[aps,floats,floatfix,superscriptaddress]{revtex4}

\usepackage{amsfonts,amsmath,amssymb,hyperref}
\usepackage[pdftex]{graphicx}

\newcommand{\arccosh}{\mathrm{arccosh}}
\newcommand{\arcsinh}{\,\mathrm{arcsinh}\,}
\newcommand{\arcsec}{\mathrm{arcsec}}

\newcommand{\M}{\mathbb{M}}
\renewcommand{\S}{\mathbb{S}}
\renewcommand{\H}{\mathbb{H}}
\newcommand{\R}{\mathbb{R}}
\newcommand{\T}{t_0}
\renewcommand{\leq}{\leqslant}

\begin{document}

\title{Network Cosmology}

\author{Dmitri Krioukov}
\affiliation{Cooperative Association for Internet Data Analysis (CAIDA),
University of California, San Diego (UCSD), La Jolla, CA 92093, USA}

\author{Maksim Kitsak}
\affiliation{Cooperative Association for Internet Data Analysis (CAIDA),
University of California, San Diego (UCSD), La Jolla, CA 92093, USA}

\author{Robert S.\ Sinkovits}
\affiliation{San Diego Supercomputer Center (SDSC),
University of California, San Diego (UCSD), La Jolla, CA 92093, USA}

\author{David Rideout}
\affiliation{Department of Mathematics,
University of California, San Diego (UCSD), La Jolla, CA 92093, USA}

\author{David Meyer}
\affiliation{Department of Mathematics,
University of California, San Diego (UCSD), La Jolla, CA 92093, USA}

\author{Mari{\'a}n Bogu{\~n}{\'a}}
\affiliation{Departament de F{\'\i}sica Fonamental,
Universitat de Barcelona, Mart\'{\i} i Franqu\`es 1, 08028 Barcelona, Spain}

\begin{abstract}
Prediction and control of the dynamics of complex networks is a central problem in network science.
Structural and dynamical similarities of different real networks suggest that some universal laws
might accurately describe the dynamics of these networks, albeit the nature and common origin of such laws remain elusive.
Here we show that the causal network representing the large-scale structure of spacetime in our accelerating universe
is a power-law graph with strong clustering, similar to many complex networks such as the Internet, social, or biological networks.
We prove that this structural similarity is a consequence of the asymptotic equivalence between the large-scale growth dynamics of complex networks and causal networks.
This equivalence suggests that unexpectedly similar laws govern the dynamics of complex networks and spacetime in the universe,
with implications to network science and cosmology.
\end{abstract}

\maketitle

\section{Introduction}

Physics explains complex phenomena in nature by reducing them to an interplay of simple fundamental laws.
This very successful tradition seems to experience certain difficulties in application to complex systems in general,
and to complex networks in particular, where it remains unclear if there exist some unique universal laws explaining a variety of
structural and dynamical similarities found in many different real networks~\cite{BaOl04,BuSp09,YaBo09,LaPe09,Vespignani2009,LiSl11,SiGo12}.
One could potentially remedy this situation by identifying a well-understood physical system whose large-scale dynamics would be asymptotically identical to the dynamics of complex networks. One could then try to use the extensively studied dynamical laws of that physical system to predict and possibly control the dynamics of networks. At the first glance, this programme seems to be quite difficult to execute, as there are no indications where to start. Yet we show here that there exists a very simple but completely unexpected connection between networks and cosmology.

In cosmology, de Sitter spacetime plays a central role as the exact solution of Einstein's equations for an empty universe, to which our universe asymptotically converges.
Here we show that graphs encoding the large-scale causal structure of de Sitter spacetime and our universe have structure common to many complex networks~\cite{DorMen-book03,newman03c-review,BoLaMoChHw06}, and that the large-scale growth dynamics of these causal graphs and complex networks are asymptotically the same. To show this, we describe the causal graphs first.

\begin{figure}
\centerline{\includegraphics[width=5.5in]{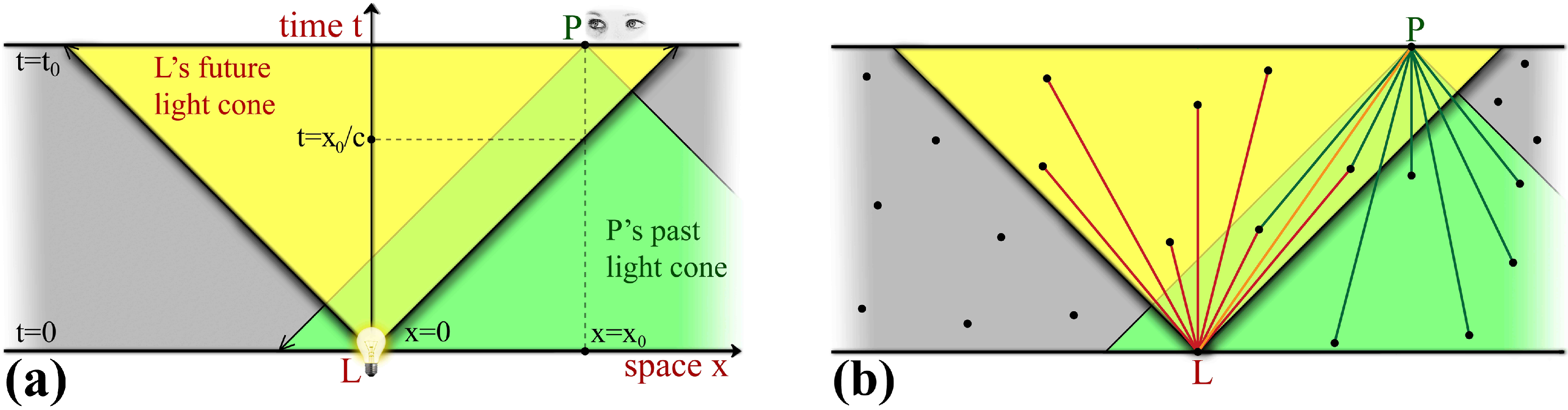}}
\caption{
Finite speed of light~$c$, and causal structure of spacetime. In panel~(a), a light source located at spatial coordinate $x=0$ is switched on at time
$t=0$. This event, denoted by $L$ in the figure, is not immediately visible to an observer located at distance $x_0$
from the light source. The observer does not see any light until time $t=x_0/c$. Since no signal can propagate faster
than $c$, the events on the observer's world line, shown by the vertical dashed line, are not causally related to $L$
until the world line enters the $L$'s future light cone (yellow color) at $t=x_0/c$. This light cone depicts the set of
events that $L$ can causally influence. An example is event $P$ located on the observer's world line $x=x_0$ at time
$t=\T>x_0/c$. The past light cone of $P$ (green color) is the set of events that can causally influence $P$. Events $L$
and $P$ lie within each other's light cones. Panel~(b) shows a set of points sprinkled into the considered spacetime patch.
The red and green links show all causal connections of events $L$ and $P$ in the resulting causet. These links form a subset
of all the links in the causet (not shown).
\label{fig:1}}
\end{figure}

The finite speed of light~$c$ is a fundamental constant of our physical world, responsible for the non-trivial causal structure of the universe~\cite{HawkingEllis1975}. If in some coordinate system the spatial distance~$x$ between two spacetime events (points in space and time) is larger than~$ct$, where $t$ is the time difference between them, then these two events cannot be causally related since no signal can propagate faster than~$c$~(Fig.~\ref{fig:1}(a)). Causality is fundamental not only in physics, but also in fields as disparate as distributed systems~\cite{Mattern88,KoNo03}
and philosophy~\cite{Causality2008}.

The main physical motivation for quantum gravity is that at the Planck scale ($l_P \sim 10^{-35}$ meters and $t_P \sim 10^{-43}$ seconds),
one expects spacetime not to be continuous but to have a discrete structure~\cite{Kiefer2007}, similar to
ordinary matter, which is not continuous at atomic scales but instead is composed of discrete atoms.
The mathematical fact that the structure of a relativistic spacetime is almost fully determined by its causal
structure alone~\cite{Malament77,HaKi76,Zeeman64} motivates the causal set approach to quantum gravity~\cite{BoLe87}.
This approach
postulates that spacetime at the Planck scale is a discrete causal set, or {\em causet}. A causet is a set of
elements (Planck-scale ``atoms'' of spacetime) endowed with causal relationships among them. A causet is thus a network in which
nodes are spacetime quanta, and links are causal relationships between them.
To make contact with General Relativity, one expects the theory to give rise to causal sets which
are constructed by a Poisson process, i.e.\ by sprinkling points into spacetime uniformly at random,
and then connecting each pair of points iff they lie within each other's light cones~(Fig.~\ref{fig:1}(b)).
According to the theorem in~\cite{BoHe09}, causets constructed by Poisson sprinkling are relativistically invariant,
as opposed to regular lattices, for example. Therefore we will use Poisson sprinkling here
to construct causets corresponding to spacetimes.
An important goal in causal set quantum gravity (not discussed here) is to identify
fundamental physical laws of causet growth consistent with Poisson sprinkling onto realistic spacetimes in the classical limit~\cite{RiSo99,AhRi10}.

In 1998 the expansion of our universe was found to be accelerating~\cite{Perlmutter98,Reiss98}. Positive vacuum energy,
or {\em dark energy}, corresponding to a positive cosmological constant $\Lambda$ in the Einstein
equations~(\ref{eq:einstein}),
is currently the most plausible explanation for this acceleration, even though the origin and nature of dark energy
is one of the deepest mysteries in contemporary science~\cite{Albrecht2006}.
Positive $\Lambda$ implies that the universe is asymptotically (at late times)
described by de Sitter spacetime~\cite{GriffithsPodolsky2009,Galloway2007}.
We first consider the structure of causets sprinkled onto de Sitter spacetime, then quantify how different this structure is for the real universe, and finally prove the asymptotic equivalence between the growth dynamics of de Sitter causets and complex networks.

\section{Results}

\subsection{Structure of de Sitter causets}

\begin{figure}
\centerline{\includegraphics[width=5.5in]{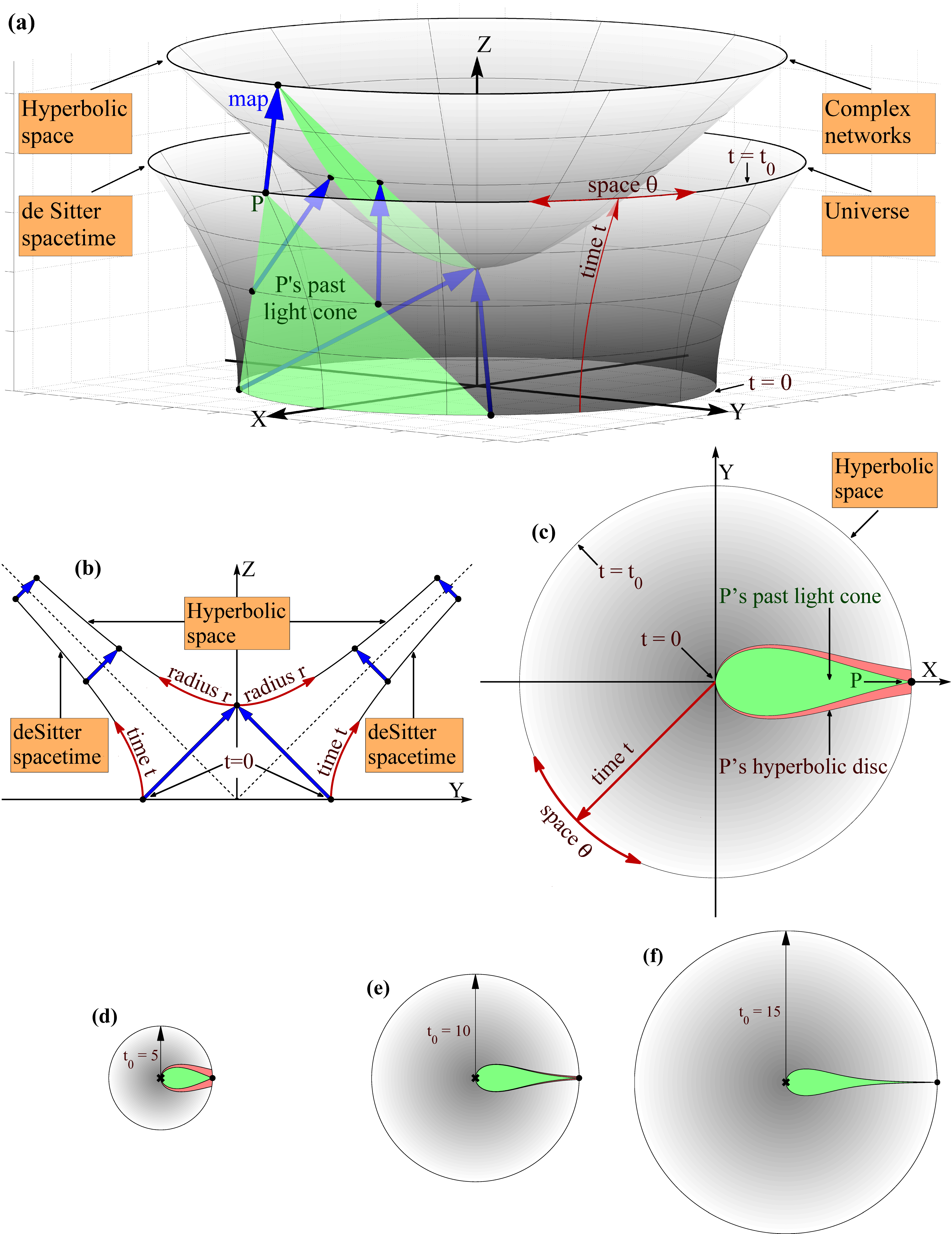}}
\caption{
Mapping between the de Sitter universe and complex
networks. Panel~(a) shows the $1+1$-dimensional de Sitter spacetime
represented by the upper half of the outer one-sheeted hyperboloid in the
$3$-dimensional Minkowski space~$XYZ$.
The spacetime coordinates $(\theta,t)$, shown by the red arrows, cover the
whole de Sitter spacetime. The spatial coordinate~$\theta_0$ of any spacetime
event, e.g.\ point~$P$, is its polar angle in the $XY$ plane, while $P$'s temporal coordinate~$t_0$ is the length
of the arc lying on the hyperboloid and connecting the point to the $XY$ plane where $t=0$. At any
time~$t$, the spatial slice of the spacetime is a circle. This
$1$-dimensional space expands exponentially with time. Dual to the outer
hyperboloid is the inner hyperboloid---the hyperbolic $2$-dimensional space,
i.e.\ the hyperbolic plane, represented by the upper sheet of a two-sheeted
hyperboloid.
The mapping between the two hyperboloids is
shown by the blue arrows. The green shapes show the past light cone of
point~$P$ in the de Sitter spacetime, and the projection of this light cone onto
the hyperbolic plane under the mapping.
Panel~(b) depicts the
cut of panel~(a) by the $YZ$ plane to further illustrate the mapping, shown
also by the blue arrows. The mapping is the reflection between the two
hyperboloids with respect to the cone shown by the dashed lines. Panel~(c)
projects the inner hyperboloid (the hyperbolic plane) with the $P$'s past
light cone (the green shape) onto the $XY$ plane. The red shape is the left
half of the hyperbolic disc centered at $P$ and having the radius equal to
$P$'s time~$t_0$, which in this representation is $P$'s radial coordinate,
i.e.\ the distance between $P$ and the origin of the $XY$ plane.
The green and red shapes become
indistinguishable at large times~$t_0$ as shown in panels~(d,e,f)
where these shapes are drawn for $t_0=5,10,15$ using the exact expressions
from Section~\ref{sec:theory}.
Assuming the average degree of $\bar{k}=10$, these $t_0$ times correspond to network sizes of approximately $40$, $200$, and $2000$ nodes.
\label{fig:2}}
\end{figure}

De Sitter spacetime is the solution of Einstein's field equations for an empty universe with positive cosmological constant $\Lambda$.
The $1+1$-dimensional de Sitter spacetime (the first~`$1$' stands for the space dimension; the second~`$1$'---for time) can be
visualized as a one-sheeted $2$-dimensional hyperboloid embedded in a flat $3$-dimensional Minkowski space~(Fig.~\ref{fig:2}(a)).
The length of horizontal circles in Fig.~\ref{fig:2}(a),
corresponding to the volume of space at a moment of time, grows
exponentially with time~$t$. Since causet nodes are distributed uniformly
over spacetime, their number also grows exponentially with time,
while as we show below, their degree decays exponentially, resulting in a power-law degree distribution in the causet.

To obtain this result, we consider in Fig.~\ref{fig:2}(a) a patch of $1+1$-dimensional de Sitter spacetime between times
$t=0$ (the ``big bang'') and $t=\T>0$ (the ``current'' time), and sprinkle $N$ nodes onto it with uniform density~$\delta$. In this spacetime the element of length $ds$ (often called {\it the metric} because its expression contains the full information about the metric tensor) and volume $dV$ (or area, since the spacetime is two-dimensional) are given
by the following expressions (Section~\ref{sec:de-sitter}):
\begin{eqnarray}
ds^2 &=& -dt^2 + \cosh^2t\,d\theta^2,\label{eq:metric-ds2}\\
dV &=& \cosh t\, dt\, d\theta,
\end{eqnarray}
where $\theta\in[0,2\pi)$ is the angular (space) coordinate on the hyperboloid. In view of the last equation and uniform sprinkling, implying that the expected number of nodes $dN$ in spacetime volume $dV$ is $dN=\delta\,dV$, the temporal node density $\rho(t)$ at time $t\in[0,\T]$ is
\begin{equation}\label{eq:rho(t)}
\rho(t)=\frac{\cosh t}{\sinh \T}\approx e^{t-\T},
\end{equation}
where the last approximation holds for $\T>t\gg1$.

Since links between two nodes in the causet exist only if the nodes lie within each other's light cones, the expected degree $\bar{k}(t)$ of a node at time coordinate $t\in[0,\T]$ is proportional to the sum of the volumes of two light cones centered at the node: the past light cone cut below at $t=0$, and the future light cone cut above at $t=\T$, similar to Fig.~\ref{fig:1}. Denoting these volumes by $V_p(t)$ and $V_f(t)$, and orienting causet links from the future to the past, i.e.\ from nodes with higher $t$ to nodes with lower $t$, we can write $\bar{k}(t)=\bar{k}_o(t)+\bar{k}_i(t)$, where $\bar{k}_o(t)=\delta V_p(t)$ and $\bar{k}_i(t)=\delta V_f(t)$ are the expected out- and in-degrees of the node. One way to compute $V_p(t)$ and $V_f(t)$ is to calculate the expressions for the light cone boundaries in the $(t,\theta)$ coordinates, and then integrate the volume form $dV$ within these boundaries. An easier way is to switch from cosmological time $t$ to conformal time~\cite{GriffithsPodolsky2009} $\eta(t)=\arcsec \cosh t$. After this coordinate change, the metric becomes conformally flat, i.e.\ proportional to the metric $ds^2=-dt^2+dx^2$ in the flat Minkowski space in Fig.~\ref{fig:1},
\begin{eqnarray}
ds^2 &=& \sec^2\eta\left(-d\eta^2 + d\theta^2\right),\\
dV &=& \sec^2\eta\,d\eta\,d\theta,
\end{eqnarray}
so that the light cone boundaries are straight lines intersecting the coordinate $(\eta,\theta)$-axes at~$45^o$, as in Fig.~\ref{fig:1} with $(t,x)$ replaced by $(\eta,\theta)$. Therefore, the volumes can be easily calculated:
\begin{eqnarray}
V_p(t) &=& \int_{0}^{\eta(t)}d\eta'\int_0^{\eta(t)-\eta'}d\theta\,\sec^2\eta'=\ln\sec\eta(t)=\ln\cosh t \approx t,\\
V_f(t) &=& \int_{\eta(t)}^{\eta(\T)}d\eta'\int_0^{\eta'-\eta(t)}d\theta\,\sec^2\eta'=\left[\eta(\T)-\eta(t)\right]\sinh\T + \ln\frac{\cosh t}{\cosh \T} \approx e^{\T-t},
\end{eqnarray}
where approximations hold for $\T\gg t\gg1$, and where we have used $\eta(t)=\arcsec\cosh t \approx \pi/2-2e^{-t}$.
For large times $t\gg1$, the past volume and consequently the out-degree are negligible compared to the future volume and in-degree, which decay exponentially with time $t$,
\begin{equation}\label{eq:k(t)}
\bar{k}_i(t) = \delta V_f(t) \approx \delta e^{\T-t}.
\end{equation}

These results can be generalized to $d+1$-dimensional de Sitter spacetimes with any $d$ and any curvature $K=\Lambda/3=1/a^2$, where $a$, the inverse square root of curvature, is also known as the curvature radius of the de Sitter hyperboloid, or as its pseudoradius. Generalizing Eqs.~(\ref{eq:rho(t)},\ref{eq:k(t)}), we can show that the temporal density of nodes and their expected in-degree in this case scale as $e^{\alpha(t-\T)}$ and $e^{\beta(\T-t)}$ with $\alpha=\beta=d/a$.
In short, we have a combination of two exponentials, number of nodes $\sim e^{\alpha t}$ born at time~$t$ and their degrees $\sim e^{-\beta t}$.
This combination yields a power-law distribution $P(k)\sim k^{-\gamma}$ of node degrees $k$ in the causet, where exponent $\gamma=1+\alpha/\beta=2$.

\begin{figure}
\centerline{\includegraphics[width=5.5in]{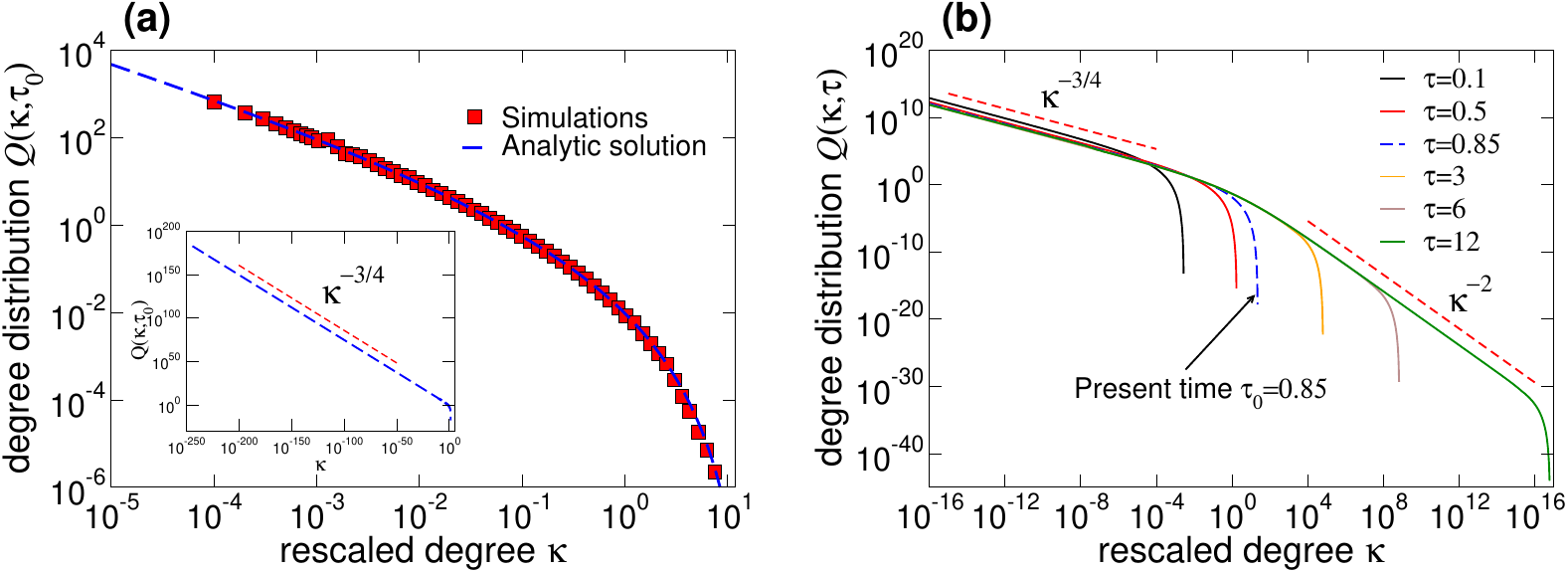}}
\caption{
Degree distribution in the universe. Panel~(a) shows
the rescaled distribution $Q(\kappa,\tau_0)=\delta a^4 P(k,\T)$ of rescaled
degrees $\kappa=k/(\delta a^4)$ in the universe causet at the present rescaled time
$\tau_0=\T/a=0.85$, where $\delta$ is the constant node density in spacetime, and $a=\sqrt{3/\Lambda}$. As shown
in Section~\ref{sec:universe},
the rescaled degree distribution
does not depend on either $\delta$ or $a$, so we set them to $\delta=10^4$
and $a=1$ for convenience. The size $N$ of simulated causets can be also set to any value without affecting the degree distribution, and this value is $N=10^6$ nodes in the figure. The degree distribution in this simulated causet is juxtaposed against the numeric evaluation of the analytical solution for $Q(\kappa,\tau_0)$ shown by the blue dashed line. The inset shows this analytic solution for the whole range of node degrees $k\in[1,10^{244}]$ in the universe, where $\delta\sim10^{173}$ and $a\sim5\times10^{17}$. Panel~(b) shows the same solution for different values of the present rescaled time $\tau$, tracing the evolution of the degree distribution in the universe in its past and future.
All further details are
in Sections~\ref{sec:universe-simulations},\ref{sec:universe}.
\label{fig:3}}
\end{figure}

\subsection{Structure of the universe and complex networks}

The large-scale causet structure of the universe in the standard model differs from the structure of sparse de Sitter causets in many ways, two of which are particularly important. First, the universe is not empty but contains matter. Therefore it is only {\em asymptotically\/} de Sitter~\cite{GriffithsPodolsky2009,Galloway2007}, meaning that only at large times $t \gg a$, or rescaled times $\tau\equiv t/a\gg1$, space in the universe expands asymptotically the same way as in de Sitter spacetime. In a homogeneous and isotropic universe, the metric is $ds^2 = -dt^2 + R^2(t)\,d\Omega^2$, where $d\Omega$ is the spatial part of the metric, and function $R(t)$ is called the scale factor. In de Sitter spacetime, the scale factor is $R(t)\sim\cosh\tau$, while in a flat universe containing only matter and dark energy, $R(t)\sim\sinh^{2/3}\left(3\tau/2\right)$. In both cases, $R(t)\sim e^\tau$ at large times $\tau\gg1$, but at early times $\tau\lesssim1$ the scaling is different. In particular, at $\tau\to0$ the universe scale factor goes to zero, resulting in a real big bang. The second difference is even more important: the product between the square of inverse curvature $a^4=1/K^2$ and sprinkling density $\delta=1/(l_P^3t_P)$ (one causet element per unit Planck $4$-volume) is astronomically huge in the universe, $\delta a^4 \sim 10^{244}$, compared to $\delta a^{d+1}\lesssim1$ in sparse causets with a small average degree. Collectively these two differences result in that the present universe causet is also a power-law graph, but with a different exponent $\gamma=3/4$ (Fig.~\ref{fig:3}(a)).

However, the $\gamma=2$ scaling currently emerges (Fig.~\ref{fig:3}(b))
as a part of a cosmic coincidence known as the ``{\em why now?}'' puzzle~\cite{GaLi99,So07,BaSh11,HaSh12}.
The matter and dark energy densities happen to be of the same order
of magnitude in the universe today.
This coincidence implies that the current rescaled time $\tau_0\equiv\T/a$ is approximately~$1$.
Figure~\ref{fig:3}(b) traces the evolution of the degree distribution
in the universe in its past and future. In the matter-dominated era with $\tau<1$,
the degree distribution is a power law with exponent $3/4$ up to a soft cut-off that grows with time. Above this
soft cut-off, the distribution decays sharply. Once we reach times $\tau\sim1$, e.g.\ today, we enter the
dark-energy-dominated era. The part of the distribution with exponent $3/4$
freezes, while the soft cut-off transforms
into a crossover to another power law with exponent $2$, whose cut-off grows exponentially with time. The crossover
point is located at $k_{cr}\sim\delta a^4$.
Nodes of small degrees $k<k_{cr}$ obey the $\gamma=3/4$ part of the distribution, while high-degree nodes, $k>k_{cr}$, lie in
its $\gamma=2$ regime. At the future infinity $\tau\to\infty$, the distribution becomes a perfect double power law with
exponents $3/4$ and $2$.

\begin{figure}
\centerline{\includegraphics[width=5.5in]{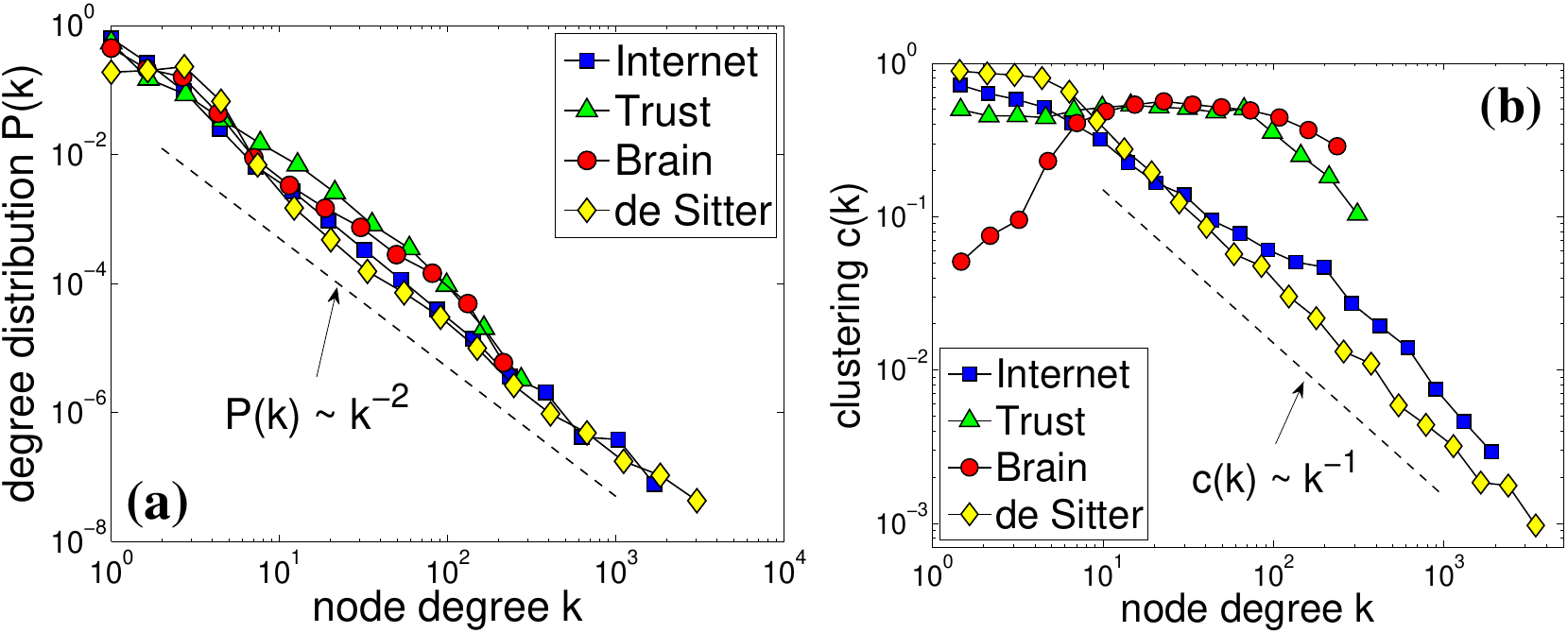}}
\caption{
Degree distribution and clustering in complex networks and de Sitter spacetime. The {\it Internet\/} is the network representing economic relations between autonomous systems, extracted from CAIDA's Internet topology measurements~\cite{ClHy09}. The network size is $N=23752$ nodes, average degree $\bar{k}=4.92$, and average clustering $\bar{c}=0.61$. {\it Trust\/} is the social network of trust relations between people extracted from the Pretty Good Privacy (PGP) data~\cite{pgp}; $N=23797$, $\bar{k}=7.86$, $\bar{c}=0.48$. {\it Brain\/} is the functional network of the human brain obtained from the fMRI measurements in~\cite{EgCh05}; $N=23713$, $\bar{k}=6.14$, $\bar{c}=0.16$. {\it De Sitter} is a causal set in the $1+1$-dimensional de Sitter spacetime; $N=23739$, $\bar{k}=5.65$, $\bar{c}=0.82$. Panel~(a) shows the degree distribution $P(k)$, i.e.\ the number of nodes $N(k)$ of degree~$k$ divided by the total number of nodes~$N$ in the networks, $P(k)=N(k)/N$, so that $\bar{k}=\sum_kkP(k)$. Panel~(b) shows average clustering of degree-$k$ nodes $c(k)$, i.e.\ the number of triangular subgraphs containing nodes of degree~$k$, divided by $N(k)k(k-1)/2$, so that $\bar{c}=\sum_kc(k)P(k)$.
All further details are
in Sections~\ref{sec:internet}-\ref{sec:desitter-simulations}.
\label{fig:4}}
\end{figure}

In short, the main structural property of the causet in the present-day universe is that it is a graph with a power-law degree distribution,
which currently transitions from the past matter-dominated era ($\tau<1$) with exponent $\gamma=3/4$ to the future dark-energy-dominated era ($\tau>1$) with $\gamma=2$.
In many (but not all) complex networks the degree distribution is also a power law with $\gamma$ close to~$2$~\cite{DorMen-book03,newman03c-review,BoLaMoChHw06}.
In Fig.~\ref{fig:4}(a) we show a few paradigmatic examples of large-scale technological, social, and biological networks for which reliable data are available, and juxtapose these networks against a de Sitter causet. In all the shown networks, the exponent $\gamma\approx2$.
This does not mean however that the networks are the same in all other respects. Degree-dependent clustering, for example (Fig.~\ref{fig:4}(b)), is
different in different networks, although average clustering is strong in all the networks. Strong clustering is another structural property often observed in complex networks:
average clustering in random graphs of similar size and average degree is lower by orders of magnitude~\cite{DorMen-book03,newman03c-review,BoLaMoChHw06}.

\subsection{Dynamics of de Sitter causets and complex networks}

Is there a connection revealing a \emph{mechanism} responsible for the emergence of this structural similarity?
Remarkably, the answer is yes.
This mechanism is the optimization of trade-offs between popularity and similarity,
shown to accurately describe the large-scale structure and dynamics of some complex networks, such as the Internet, social trust network, etc.~\cite{PaBoKr11}
The following model of growing networks, with all the parameters set to their default values, formalizes this optimization in~\cite{PaBoKr11}.
New nodes $n$ in a modeled network are born one at a time, $n=1,2,3,\ldots$, so that $n$ can be called a network time. Each new node is placed uniformly at random on circle $\S^1$. That is, the angular coordinates $\theta_n$ for new nodes $n$ are drawn from the uniform distribution on $[0,2\pi]$. Circle $\S^1$ models a similarity space. The closer the two nodes on $\S^1$, the more similar they are. All other things equal, the older the node, the more popular it is, the higher its degree. Therefore birth time $n$ of node $n$ models its popularity. Upon its birth, new node $n_0$ optimizes between popularity and similarity by establishing its fixed number $m$ of connections to $m$ existing nodes $n<n_0$ that have the minimal values of the product $n\Delta\theta$, where $\Delta\theta=\pi-|\pi-|\theta_n-\theta_{n_0}||$ is the angular distance between nodes $n$ and $n_0$. One dimension of this trade-off optimization strategy is to connect to nodes with smaller birth times $n$ (more popular nodes); the other dimension is to connect to nodes at smaller angular distances $\Delta\theta$ (more similar nodes). After placing each node $n$ at radial coordinate $r_n=\ln n$, all nodes are located on a two-dimensional plane at polar coordinates $(r_n,\theta_n)$. For each new node $n_0$, the set of nodes minimizing $n\Delta\theta$ is identical to the set of nodes minimizing $x=\ln(nn_0\Delta\theta/2)=r_n+r_{n_0}+\ln(\Delta\theta/2)$, where $x$ is equal to the hyperbolic distance~\cite{Bonahon09-book} between nodes $n_0$ and $n$ if $r$, $r_0$, and $\Delta\theta$ are sufficiently large. One can compute the expected distance from node $n_0$ to its $m$th closest node, and find that this distance is equal to $\ln[\pi mn_0/\{2(1-1/n_0)\}] \approx r_{n_0}+\ln[\pi m/2] \approx r_{n_0}$, where approximations hold for large $n_0$. In other words, new node $n_0$ is born at a random location on the edge of an expanding hyperbolic disc of radius $r_0=\ln n_0$, and connects to asymptotically all the existing nodes lying within hyperbolic distance $r_0$ from itself. The connectivity perimeter of new node $n_0$ at time $n_0$ is thus the hyperbolic disc of radius $r_0$ centered at node $n_0$.
The resulting connection condition $x<r_0$, satisfied by nodes $n$ to which new node $n_0$ connects, can be rewritten as
\begin{eqnarray}
r_n+\ln\frac{\Delta\theta}{2}<0\label{eq:ball-ineq-approx-mt}.
\end{eqnarray}
This model yields growing networks with power-law degree distribution $P(k)\sim k^{-\gamma}$ and $\gamma=2$. The networks in the model also have strongest possible clustering, i.e.\ the largest possible number of triangular subgraphs, for graphs with this degree distribution. The model and its extensions describe the large-scale structure and growth dynamics of different real networks with a remarkable accuracy~\cite{PaBoKr11}.
We next show that the described network growth dynamics is asymptotically identical to the growth dynamics of de Sitter causets.

To show this,
consider a new spacetime quantum~$P$ that has just been born at current time $t=\T$ in Fig.~\ref{fig:2}.
That is, assume that the whole de Sitter spacetime is sprinkled by nodes with a uniform density,
but only nodes between $t=0$ and $t=\T$ are considered to be ``alive.'' We can then model
causet growth as moving forward the current time boundary $t=\T$ one causet element $P$ at a time.
By the causet definition, upon its birth, $P$ connects to all nodes
in its past light cone shown by green.
As illustrated in Fig.~\ref{fig:2}, we then map the upper half of the outer one-sheeted hyperboloid representing the half of de Sitter spacetime $d\S^2$ with $t>0$,
to the upper sheet of the dual two-sheeted inner hyperboloid, which is the standard hyperboloid representation of the hyperbolic space $\H^2$~\cite{CaFlo97}.
This mapping sends a point with coordinates $(t,\theta)$ in $d\S^2$ to the point with coordinate $(r,\theta)$ in $\H^2$, where $r=t$. Since in the conformal
time coordinates the light cone boundaries are straight lines intersecting the $(\eta,\theta)$-axes at $45^o$, the coordinates $(t,\theta)$ of all points in $P$'s past light
cone satisfy inequality $\Delta\theta<\Delta\eta=\eta(\T)-\eta(t)=\arcsec\cosh\T - \arcsec\cosh t \approx 2\left(e^{-t}-e^{-\T}\right)$.
If $\T \gg t \gg 1$, then we can neglect the second term in the last expression,
and the coordinates $(t_n,\theta_n)$ of existing
causet nodes $n$ to which new node $P$ connects
upon its birth are given by
\begin{eqnarray}
\Delta\theta<2e^{-t_n},
\end{eqnarray}
which is identical to Eq.~(\ref{eq:ball-ineq-approx-mt}) since $r_n=t_n$.
In Section~\ref{sec:theory}
we fill in further details of this proof, extend it to any dimension and curvature, and show that the considered mapping between de Sitter spacetime and hyperbolic space is relativistically invariant.

In short, past light cones of new nodes, shown by green in Fig.~\ref{fig:2}, are asymptotically equal (Fig.~\ref{fig:2}(d-f)) to the corresponding hyperbolic discs, shown by red.
The green light cone bounds the set of nodes to which node $P$ connects as a new causet element.
The red hyperbolic disc bounds the set of nodes to which $P$ connects as a new node in the hyperbolic network model that accurately describes the growth of real networks.
Since these two sets are asymptotically the same, we conclude that not only the structure, but also the growth dynamics of complex networks and de Sitter causets are asymptotically identical.

\section{Discussion}

Geometrically, this equivalence is due to a simple duality between the two hyperboloids in Fig.~\ref{fig:2}.
The inner hyperboloid represents the popularity$\times$similarity hyperbolic geometry of complex networks; the outer hyperboloid is the de
Sitter spacetime, which is the solution to Einstein's equations for a universe with positive vacuum energy.
In that sense, Einstein's equations provide an adequate baseline description for the structure and dynamics of complex networks,
which can be used for predicting network dynamics at the {\em large scale}.
De Sitter spacetime is homogeneous and isotropic, as is the hyperbolic space,
but if we take a real network, e.g.\ the Internet, and map it to this homogeneous space,
then after the mapping, the node density in the space is non-uniform~\cite{BoPa10}. In real networks,
the space thus  appears homogeneous only at the largest scale, while at {\em smaller
scales\/} there are inhomogeneities and anisotropies, similar to the real universe, in which
matter introduces spacetime inhomogeneities at smaller
scales, and leads to non-trivial coupled dynamics of
matter density and spacetime curvature, described by the same Einstein equations.
In view of this analogy, equations similar to Einstein's equations may also apply to complex networks at smaller scales.
If so, these equations can be used to predict and possibly control the {\em fine-grained\/} dynamics of links
and nodes in networks.

Our results may also have important implications for cosmology. In particular, de Sitter causal sets have exactly
the same graph structure that maximizes network navigability~\cite{BoKrKc08}. Translated to asymptotically sparse causal sets, does this property imply
that the expanding portion of de Sitter spacetime ($t>0$) is the spacetime that maximizes the probability that two random Planck-scale events have an
ancestor in their common past? If it does, then this uniqueness of de Sitter spacetime may lead to a different perspective on the cosmic coincidence problem,
as well as on dark energy, possibly casting the latter as a phenomenon emerging from certain optimization principles encoded in the causal network structure.

The degree distributions in some complex networks deviate from clean power laws, the exponents of these power laws vary a lot across different real networks, and so do clustering, correlation, and many other structural properties of these networks~\cite{DorMen-book03,newman03c-review,BoLaMoChHw06}. Therefore it may seem unlikely that de Sitter causets can model the full spectrum of structural diversity observed in complex networks. Focusing on the trust network in Fig.~\ref{fig:4} for instance, we have already observed that its degree-dependent clustering is quite different from the one in de Sitter causets. Yet, given that these causets are asymptotically identical to growing hyperbolic networks, this observation appears as a paradox, because the hyperbolic networks were shown to accurately match not only clustering peculiarities, but also a long list of other important structural properties of the same trust network, as well as of other networks~\cite{PaBoKr11}. The explanation of this paradox lies in that the hyperbolic network model has parameters to tune the degree distribution exponent, clustering strength, node fitness, and other network properties, while in de Sitter causets, only the number of nodes and average degree can be controlled. Do the hyperbolic network parameters have their duals in the de Sitter settings, what are the physical meanings of these dual parameters, and do they lead to similar modeling versatility---all these questions are open.

We conclude with the observation that the node density in growing hyperbolic networks with the default parameters corresponding to de Sitter causets, is not uniform in the hyperbolic space~\cite{PaBoKr11}. This observation means that these networks are not random geometric graphs~\cite{Penrose03-book}, and that their structure does not exactly reflect the geometry of the underlying hyperbolic space. Informally, a random geometric graph is a coarse, discrete representation of a smooth geometric space. Our finding that asymptotically the same networks have a uniform node density in de Sitter spacetimes dual to hyperbolic spaces, strongly suggests that real networks are random geometric graphs that grow in spacetimes similar to the asymptotically de Sitter spacetime of our accelerating universe.

\begin{acknowledgments}
We thank M.\'A.~Serrano, M.~Norman, Z.~Toroczkai, A.-L.~Barab\'asi, J.~Garriga, A.~Vilenkin, R.~Sorkin, G.~Gibbons, F.~Papadopoulos, G.~Bianconi, A.~Petersen, L.~Braunstein, F.~Bonahon, kc~claffy, C.~Carlson, L.~Klushin, S.~Paston, V.D.~Lyakhovsky, and A.~Krioukov for useful discussions and suggestions. Special thanks to D.~Chialvo for sharing his brain data with us. We also thank M.~Norman for providing computing time at the SDSC through the director's discretionary fund, and to J.~Cheng and B.~Huffaker for their help with Fig.~\ref{fig:1}. This work was supported by DARPA grant No.\ HR0011-12-1-0012; NSF grants No.\ CNS-0964236 and CNS-1039646; Cisco Systems; Foundational Questions Institute grant No.\ FQXi-RFP3-1018; George W.\ and Carol A.\ Lattimer Campus Professorship at UCSD; MICINN project No.\ FIS2010-21781-C02-02; Generalitat de Catalunya grant No.\ 2009SGR838; and by the ICREA Academia prize, funded by the Generalitat de Catalunya.
\end{acknowledgments}

\appendix

\pagebreak
\section{Data and Simulations}\label{sec:data}

Here we describe the network data and methods used in Figs.~\ref{fig:3}\&\ref{fig:4}. The considered real networks are paradigmatic examples of complex technological, social, and biological networks for which reliable large-scale data are available. We note that in all the three considered networks, links represent soft relational data instead of hard-wired network diagrams: economic/business relations in the Internet, trust relations between people, causal/correlation relations in the brain, and causal relations in causal sets.

\subsection{Internet data}\label{sec:internet}

The Internet topology in Fig.~\ref{fig:4} represents economic/business relations between {\it autonomous systems} or {\it ASs}. The AS is an organization or an individual owning a part of the Internet infrastructure. The network is extracted from the data collected by CAIDA's Archipelago Measurement Infrastructure (Ark)~\cite{ClHy09}, \url{http://www.caida.org/projects/ark/}. The Ark infrastructure consists of a set of monitors continuously tracing IP-level paths to random destinations in the Internet. The union of the paths collected by all the monitors is then aggregated over a certain period of time, and each IP address in the collection is mapped to an AS owning this address, using the RouteViews BGP tables~\url{http://www.routeviews.org/}. The resulting AS network has a power-law degree distribution with exponent $\gamma=2.1$, and this result is stable over time~\cite{SiFaFaFa03,DhDo08} and across different measurement methodologies~\cite{ZhaLiMaZha05,MaKrFo06}, \url{http://www.caida.org/research/topology/topo_comparison/}. The data and its further description are available at~\url{http://www.caida.org/data/active/ipv4_routed_topology_aslinks_dataset.xml}. In Fig.~\ref{fig:4}, the data for June 2009 is used. The aggregation period is one month, and the number of monitors is~$36$.

\subsection{Trust data}

The trust network in Fig.~\ref{fig:4} represents trust relations between people, extracted from the Pretty-Good-Privacy~(PGP) data.
PGP is a data encryption and decryption computer program that provides cryptographic privacy and
authentication for data communication~\cite{pgp}, \url{http://www.openpgp.org/}. In the PGP trust network, nodes are
certificates consisting of public PGP keys and owner information. A directed link in the network pointing from
certificate {\it A} to certificate {\it B} represents a digital signature by the owner of {\it A}, endorsing the
owner/public key association of {\it B}. We use the PGP web of trust data collected and maintained by J\"{o}rgen
Cederl\"{o}f~\url{http://www.lysator.liu.se/~jc/wotsap/wots2/}. We post-process the data as follows.
The directed network is first mapped to its undirected counterpart by taking into account only bi-directional trust links between the certificates.
The largest connected component is then extracted from this undirected network. This way we strengthen the social aspect of the network, since we consider only pairs of users (owners of PGP keys) who have reciprocally signed each other's keys. Such filtering increases the probability that the
connected users know each other, and makes the extracted network a reliable proxy to the underlying social network. The degree distribution in the resulting network is a power law with exponent $\gamma=2.1$, and this result is stable over time~\cite{PaBoKr11}, \url{http://pgp.cs.uu.nl/plot/}. In Fig.~\ref{fig:4}, the data snapshot taken on December 1, 2005, is used.

\subsection{Brain data}

The brain functional network in Fig.~\ref{fig:4} represents causal/correlation relations between small areas in the human brain. This network is extracted from the fMRI measurements in~\cite{EgCh05}. In those experiments, the whole brains of different human subjects are split into $36\times64\times64=147456$ adjacent areas called {\it voxels}, each voxel of volume $3\times3.475\times3.475\text{ mm}^3$. The subjects are then asked to perform different tasks, during which the magnetic resonance activity $V(x,t)$ is recorded at each voxel $x$ at time $t$. Time $t$ is discrete: $400$ recordings are made with the interval of $2.5\text{ s}$. Given this data, and denoting by $\langle\cdot\rangle$ the time average, the correlation coefficient $r(x,x')$ for each pair of voxels is then computed:
\begin{equation}
r(x,x') = \frac{\langle V(x,t)V(x',t) \rangle - \langle V(x,t) \rangle \langle V(x',t) \rangle}
{\sqrt{\left[\langle V(x,t)^2 \rangle - \langle V(x,t) \rangle^2\right] \left[\langle V(x',t)^2 \rangle - \langle V(x',t) \rangle^2\right]}}.
\end{equation}
To form a functional network out of this correlation data, two voxels $x$ and
$x'$ are considered causally connected if the correlation coefficient between
them exceeds a certain threshold~$r_c$, $r(x,x')>r_c$. If $r_c$ is too small
or too large, then the resulting network is fully connected or fully
disconnected. There exists however a unique percolation transition value of
$r_c$ corresponding to the onset of the giant component in the network. To
find this value, we compute the sizes $|S_1|_{r_c}$ and $|S_2|_{r_c}$ of the
largest and second largest components $S_1$ and $S_2$ in the network for
different values of $r_c$. The value of threshold $r_c$ corresponding to the
largest values of $|\partial|S_1|/\partial r_c|_{r_c}$ and $|S_2|_{r_c}$ is then its percolation transition value. With threshold $r_c$ set to this value, the network has a power-law degree distribution with exponent $\gamma=2$ and exponential cutoffs, and this result is stable across different human subjects and different types of activity that they perform during measurements, see \cite{EgCh05,FrBa09} and \url{http://www.caida.org/publications/papers/2012/network_cosmology/supplemental/}. In Fig.~\ref{fig:4}, a specific dataset is used---{\it Set14}, see the URL---where the subject is at rest. The threshold value is $r_c=0.7$.

\subsection{De Sitter causet}\label{sec:desitter-simulations}

The causet in Fig.~\ref{fig:4} is generated by sprinkling a number of points over a patch in the $1+1$-dimensional de Sitter spacetime, and connecting each pair of points if they lie within the other's light cones. The point density is uniform in the de Sitter metric, and the size of the patch is such that the size of the generated causet and its average degree are close to those in the considered examples of complex networks.

In conformal time coordinates, see Section~\ref{sec:de-sitter} below, each spacetime point has two coordinates, spatial $\theta\in[0,2\pi]$ and temporal $\eta\in(-\pi/2,\pi/2)$, with $\eta=-\pi/2$ and $\eta=\pi/2$ corresponding to the past and future
infinities respectively. The spacetime patch that we consider is between $\eta=0$ and $\eta=\eta_0>0$, where $\eta_0$ is determined below. This patch is illustrated in Fig.~\ref{fig:2}(a). To sample $N$ points from this patch with uniform density, we sample $N$ pairs of random numbers: $N$ spatial coordinates $\theta$ drawn from the uniform distribution on $[0,2\pi]$, and $N$ temporal coordinates $\eta$ drawn from the distribution
\begin{equation}\label{eq:rho(eta)d=1}
\rho(\eta|\eta_0) = \frac{\sec^2\eta}{\tan\eta_0}.
\end{equation}
Two spacetime points with coordinates $(\eta,\theta)$ and $(\eta',\theta')$ are then connected in the causet if $\Delta\theta<\Delta\eta$, where $\Delta\theta=\pi-|\pi-|\theta-\theta'||$ and $\Delta\eta=|\eta-\eta'|$ are the spatial and temporal distances between the points in this coordinate system.

\begin{figure}
\centerline{\includegraphics[width=3in]{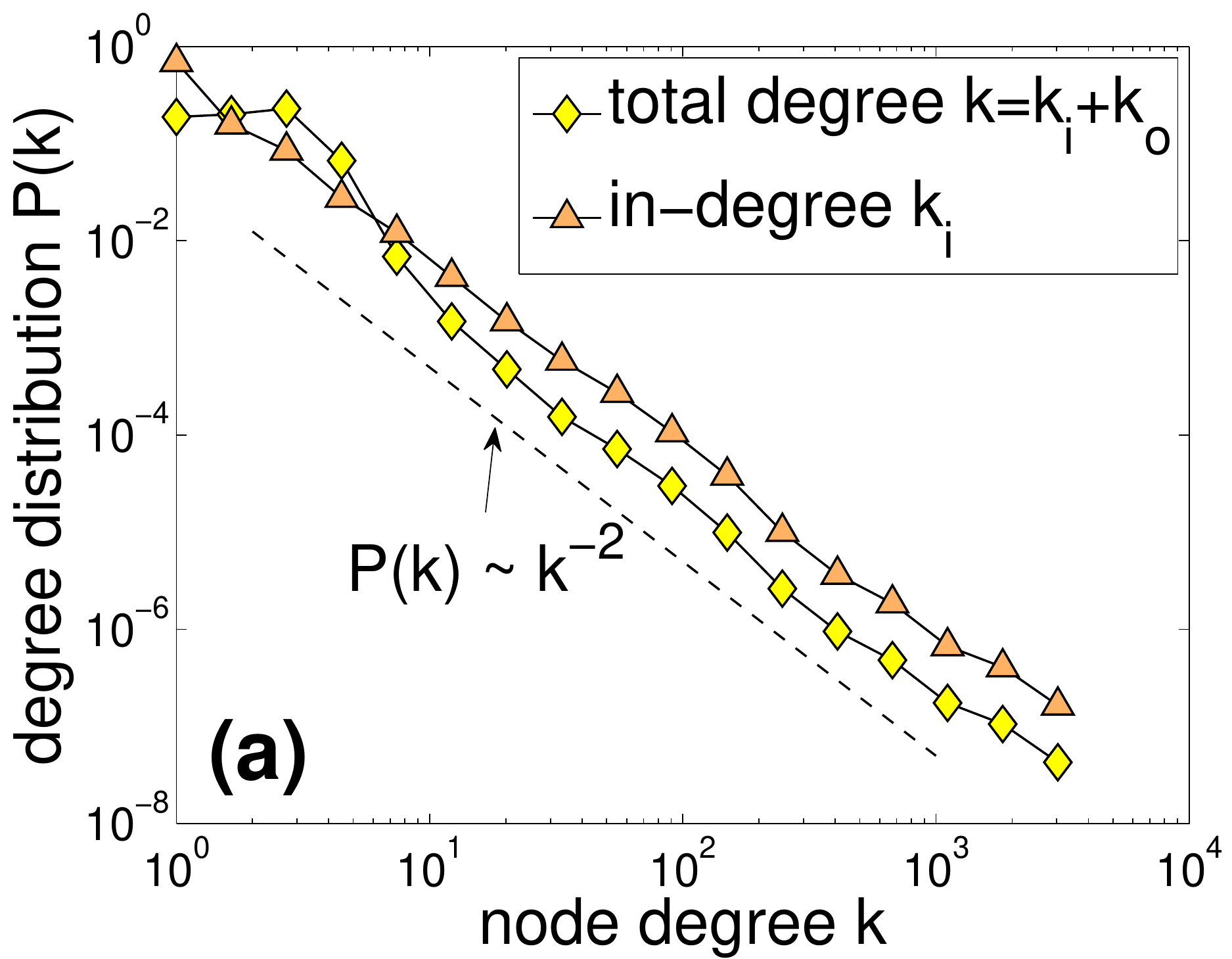}
\includegraphics[width=3in]{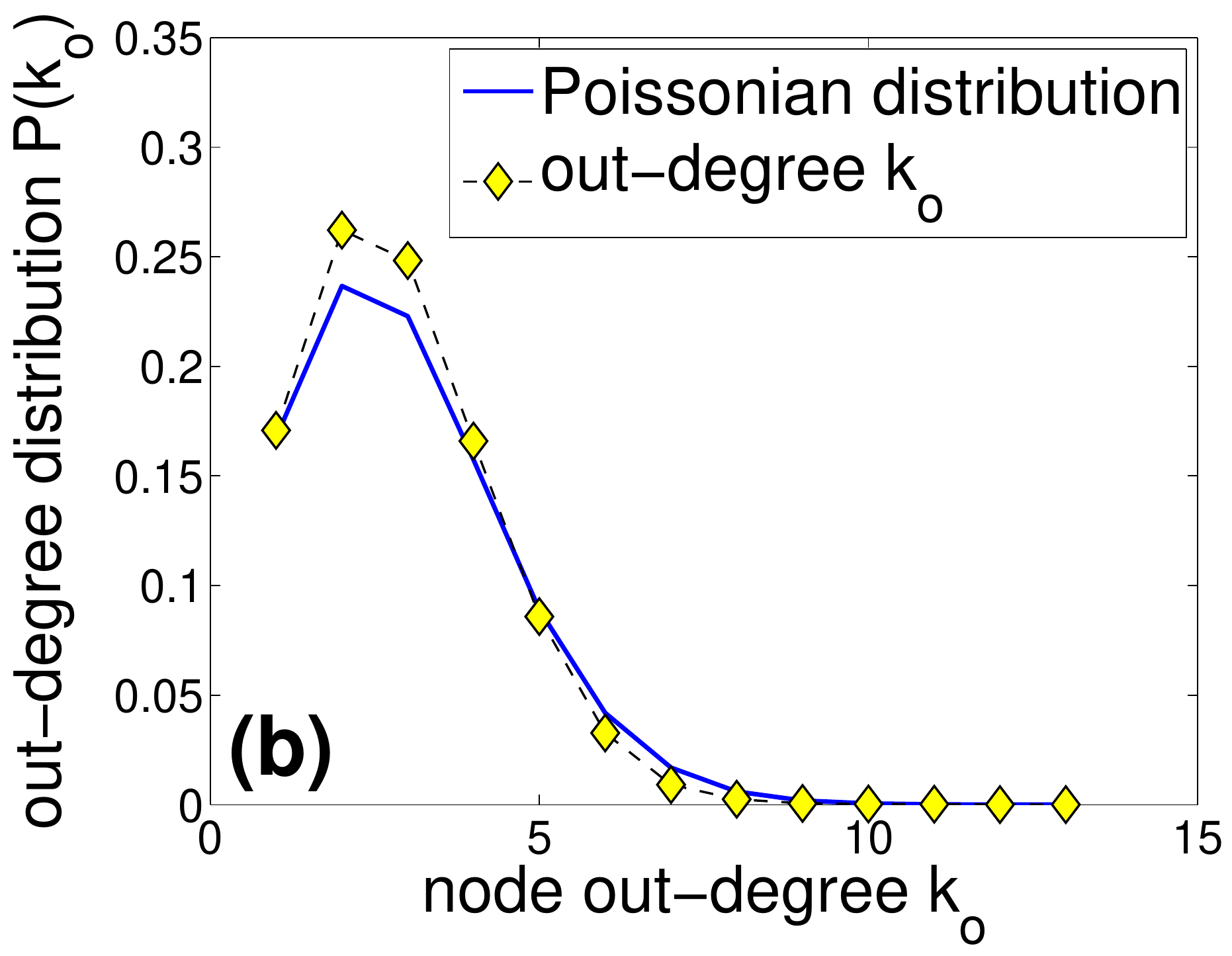}}
\caption{(a)~The total versus in-degree distributions in the de Sitter causet of Fig.~\ref{fig:4}. (b)~The out-degree distribution in the same causet. The solid line shows the Poissonian distribution with the mean $\lambda=\bar{k}_o=\bar{k}/2=2.83$, where $\bar{k}_o$ is the average out-degree in the causet.}
\label{fig:ds-sims}
\end{figure}

To determine $\eta_0$ we first note that since the point density is uniform, the number of points $N$ is proportional to the patch volume, and the proportionality coefficient is a constant point density~$\delta$. The volume of the patch is easy to calculate, see Section~\ref{sec:de-sitter}, where we can also calculate the average degree $\bar{k}$ in the resulting causets, so that we have:
\begin{eqnarray}
N &=& 2\pi\delta a^2\tan\eta_0,\\
\bar{k} &=& 4\delta a^2\left(\frac{\eta_0}{\tan\eta_0}+\ln\sec\eta_0-1\right),\\
\frac{\bar{k}}{N} &=& \frac{2}{\pi}\cdot\frac{\eta_0/\tan\eta_0+\ln\sec\eta_0-1}{\tan\eta_0},\label{eq:k/N_d=1}
\end{eqnarray}
where $a$ is the spacetime's pseudoradius determining its curvature.
Given a target average degree $\bar{k}$ and number of nodes $N$ in a causet, their ratio determines $\eta_0$ via the last equation. Sampling $N$ points from this patch will then yield causets with expected average degree $\bar{k}$.
The average degree in generated causets will not be exactly equal to $\bar{k}$ since there will be nodes of degree~$0$ and Poissonian fluctuations of the numbers of nodes lying within light cones around their expected values.
In Fig.~\ref{fig:4}, the exact number of sampled points is $N=24586$ and target value of the average degree is $\bar{k}=5.53$, so that the above equations yield $\eta_0=\pi/2-3.86\times10^{-5}$ and $\delta a^2=0.151$. The resulting number of nodes excluding nodes of degree~$0$ and their average degree in the generated causet are reported in the caption of Fig.~\ref{fig:4}.

If we direct links in the causet from the future to the past, i.e.\ from
nodes with larger $\eta$ to nodes with smaller $\eta$, then the distributions of in-degrees $k_i$ and total degrees $k=k_i+k_o$ are similar, Fig.~\ref{fig:ds-sims}(a), because the distribution of out-degrees $k_o$ is not particularly interesting, close to Poissonian, Fig.~\ref{fig:ds-sims}(b). The semantics of this link direction in causal sets is similar to that in preferential attachment, where it is often convenient to consider links oriented from new to old nodes~\cite{KraReLe00,DoMeSa00}. In preferential attachment the out-degree distribution is not particularly interesting either, and is given by the delta-function $\delta(m)$, where $m=\bar{k}/2$ is the number of connections that each new node establishes. Figure~\ref{fig:4} shows the degree distribution in the undirected causet, i.e.\ the distribution of total degrees $k=k_i+k_o$, since all the other networks shown in the figure are undirected (we consider only reciprocal trust relationships in the trust network). As a side note, the direction of links in different directed networks may have very different and network-specific semantics. In gene regulations~\cite{GuBo02} or the considered web of trust, for example, link directions show what genes regulate what other genes or who trusts whom, respectively, which may have little in common with the temporal link direction in preferential attachment or causal sets, showing what nodes are newer or older.

\begin{figure}
\centerline{\includegraphics[width=3in]{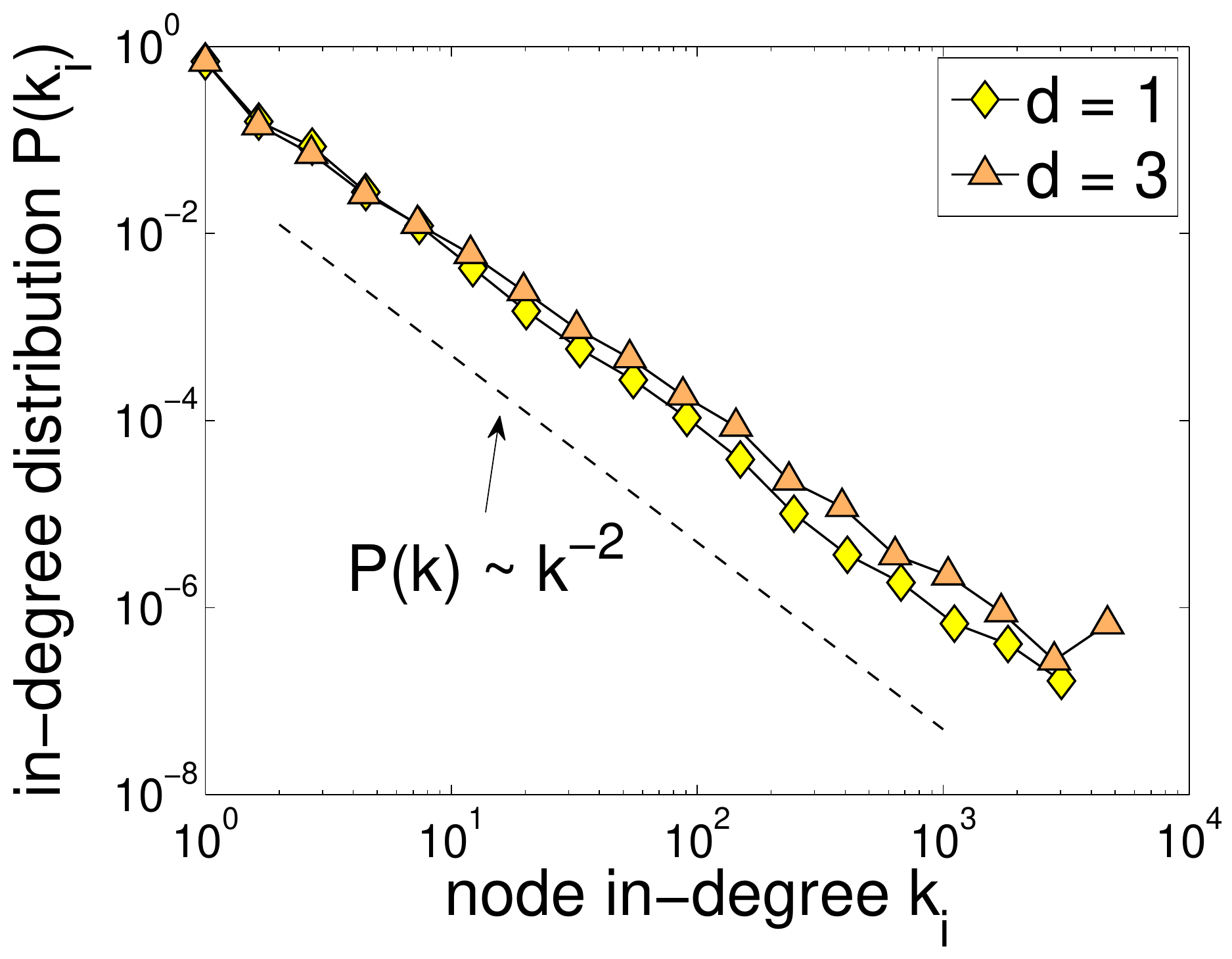}}
\caption{The in-degree distribution of causets approximated by patches of 1+1 and 3+1 dimensional de Sitter spacetime. The causet sizes are $N=23739$ and $N=23732$, and their average degrees are $\bar{k}=5.65$ and $\bar{k}=5.25$ respectively.}
\label{fig:ds-dims}
\end{figure}

In higher dimensions, pre-asymptotic effects become more prominent in causets of similar size and average degree. In particular, the exponent of the degree distribution is slightly below~$2$, see Fig.~\ref{fig:ds-dims} comparing the causets in the $d=1$ and $d=3$ cases. The former causet is the same as in Figs.~\ref{fig:4}\&\ref{fig:ds-sims}, while the latter is obtained by a procedure similar to the one described above, except that instead of Eqs.~(\ref{eq:rho(eta)d=1}-\ref{eq:k/N_d=1}), we have
\begin{eqnarray}
\rho(\eta|\eta_0) &=& \frac{3\sec^4\eta}{\left(2+\sec^2\eta_0\right)\tan\eta_0},\\
N &=& \frac{2}{3}\pi^2\delta a^4\left(2+\sec^2\eta_0\right)\tan\eta_0,\\
\bar{k} &=& \frac{4}{9}\pi\delta a^4\frac{12\left(\eta_0/\tan\eta_0+\ln\sec\eta_0\right)+\left(6\ln\sec\eta_0-5\right)\sec^2\eta_0-7}{2+\sec^2\eta_0},\\
\frac{\bar{k}}{N} &=& \frac{2}{3\pi}\cdot\frac{12\left(\eta_0/\tan\eta_0+\ln\sec\eta_0\right)+\left(6\ln\sec\eta_0-5\right)\sec^2\eta_0-7}{\left(2+\sec^2\eta_0\right)^2\tan\eta_0},
\end{eqnarray}
and each point has two additional angular coordinates, $\theta_1$ and $\theta_2$, which are random variables between $0$ and $\pi$ drawn from distributions $(2/\pi)\sin^2\theta_1$ and $(1/2)\sin\theta_2$, see Section~\ref{sec:de-sitter}. The spatial distance $\Delta\theta$ between pairs of points is then computed using the spherical law of cosines, and two points are causally linked if $\Delta\theta<\Delta\eta$ as before. In Fig.~\ref{fig:ds-dims}, the $d=3$ causet has target $N=25441$ and $\bar{k}=5.29$, yielding $\eta_0=\pi/2-4.11\times10^{-2}$ and $\delta a^4=0.267$. The maximum-likelihood fit of the degree sequence in the $d=3$ and $d=1$ causets yields $\gamma=1.65$ and $\gamma=1.90$, while the least-square fit of the complementary cumulative distribution function for node degrees yields $\gamma=1.77$ and $\gamma=1.98$.

\subsection{Simulating the universe}\label{sec:universe-simulations}

\begin{figure}
\centerline{\includegraphics[width=3.5in]{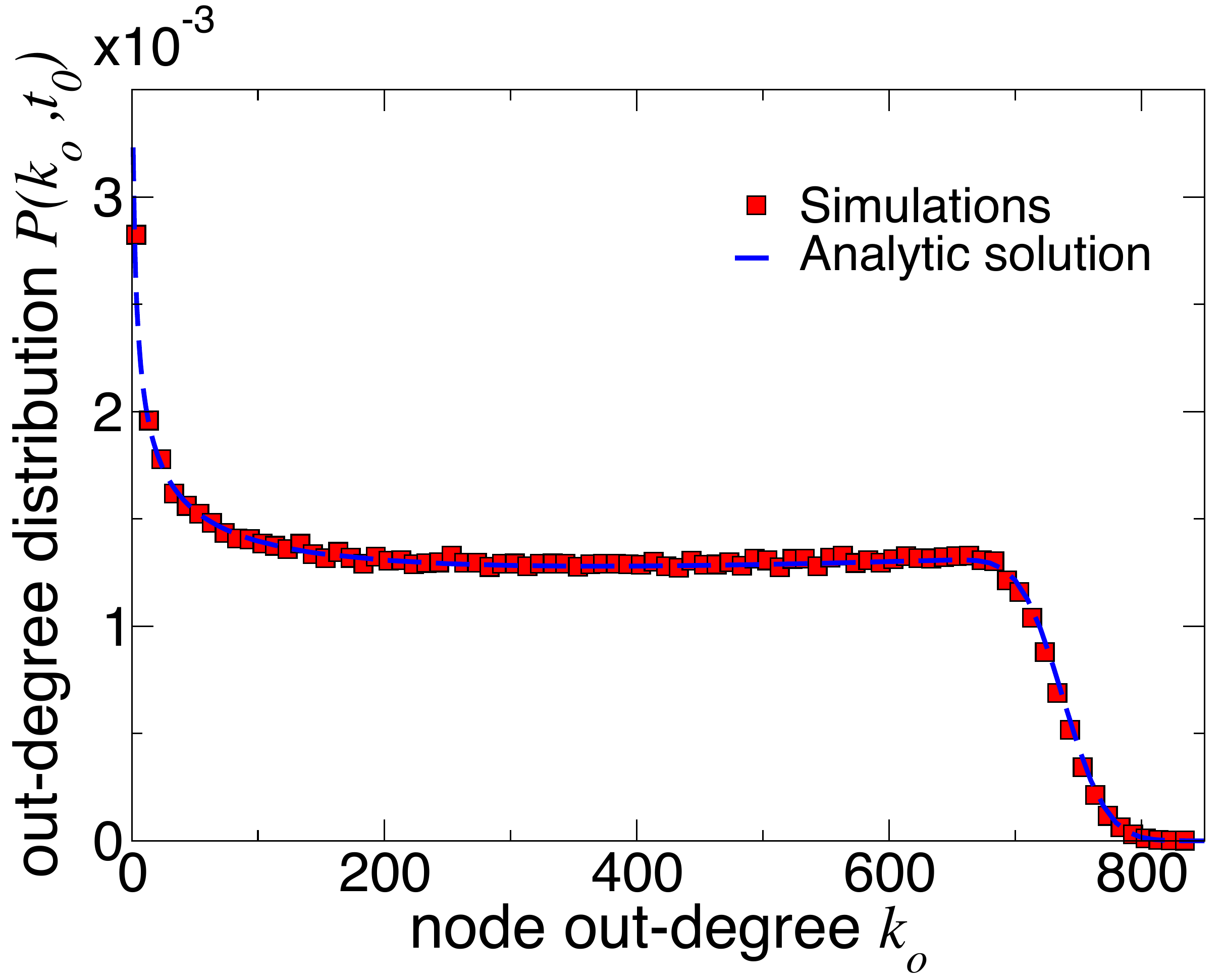}}
\caption{Out-degree distribution in the simulated universe.}
\label{fig:pkout}
\end{figure}

The results in Section~\ref{sec:universe} below allow us to simulate the universe causet similarly to the de Sitter simulations described in the previous section. The main idea behind the universe simulations is that we can use the exact results for a flat universe, but apply them to a closed universe with an arbitrary finite number of nodes in its causet, since the real universe is almost flat, and since the degree distribution in both cases is the same.

Specifically, the causet in Fig.~\ref{fig:3} is generated as follows. Unlike the previous section where only conformal time is used, here it is more convenient to begin with the rescaled time $\tau=t/a$. The current measurements of the universe yield, see Section~\ref{sec:measurements}, $\tau_0=\T/a=(2/3)\arcsinh\sqrt{\Omega_\Lambda/\Omega_M}=0.8458$ as the best estimate for this rescaled time in the universe today. According to Eqs.~(\ref{eq:scale-factor-universe},\ref{eq:sinh23}), the scale factor is
\begin{equation}\label{eq:scale-factor-sims}
R(\tau)=\alpha\sinh^{\frac{2}{3}}\frac{3}{2}\tau,
\end{equation}
where $\alpha$ is a free parameter that we can set to whatever value we wish, since the degree distribution does not depend on it, see Section~\ref{sec:scaling}. We wish to set $\alpha$ to the value such that the generated causet would have a desired number of nodes
\begin{equation}
N=\frac{1}{3}\pi^2 \delta \alpha^3\left(\sinh{3\tau_0}-3\tau_0\right),
\end{equation}
where $\delta$, the node density, is yet another free parameter that does not affect the degree distribution according to Section~\ref{sec:scaling}.
The scale factor in Eq.~(\ref{eq:scale-factor-sims}) means that the temporal coordinate that we assign to each of these $N$ nodes is a random number $\tau\in[0,\tau_0]$ drawn from distribution
\begin{equation}
\rho(\tau|\tau_0)=\frac{6 \sinh^2{\frac{3}{2}\tau}}{\sinh{3\tau_0}-3\tau_0}.
\end{equation}
Having times $\tau$ assigned, we then map them, for each node, to conformal times $\eta$ via
\begin{equation}
\eta=\frac{2}{3\alpha} \int_{0}^{\frac{3}{2}\tau} \frac{dx}{\sinh^{\frac{2}{3}}{x}}.
\end{equation}
The spatial coordinates $\theta$, $\theta_1$, and $\theta_2$ are then assigned exactly as in the previous section, and the causet network is also formed exactly the same way, i.e.\ by future$\to$past linking all node pairs whose temporal distance $\Delta\eta$ exceeds their spatial distance $\Delta\theta$.

Figure~\ref{fig:3} shows the in-degree distribution for a causet generated with $\alpha=2.01$, $a=1$, and $\delta=10^4$. The resulting number of nodes in the causet is $N=10^6$. The analytic solution curves are obtained using the approximations for the in-degree distribution derived in Section~\ref{sec:universe}. The key equations are Eq.~(\ref{eq:P(k)approx}), yielding the Laplace approximation for the in-degree distribution, and Eq.~(\ref{eq:zeta}) expressing conformal time as a function of the current value of the scale factor.
The precise steps to numerically compute the analytic solution for the in-degree distribution are listed in Section~\ref{sec:numerics}.
The perfect match between the simulations and analytic solution in Fig.~\ref{fig:3} confirms that the approximations used in Section~\ref{sec:universe} to derive the analytic solution yield very accurate results.

The out-degree distribution in the same simulated causet is shown in Fig.~\ref{fig:pkout}. The analytic solution is obtained by numeric evaluations of Eqs.(\ref{eq:ko(eta)}-\ref{eq:degreedistribution}). The distribution appears to be uniform over a wide range of degree values, which is quite different from the Poissonian out-degree distribution in the sparse de Sitter causet in Fig.~\ref{fig:ds-sims}(b) because $\delta a^{d+1} = 10^4 \gg 1$ in the universe simulations, whereas $\delta a^{d+1} < 1$ in the de Sitter simulations in the previous section. Depending on whether $\delta a^{d+1}$ is smaller or larger than~$1$, (asymptotically) de Sitter causets have either power-law in-degree distributions with $\gamma=2$ and Poissonian out-degree distributions, or double power-law in-degree distributions with $\gamma=3/4$ and $\gamma=2$, and non-trivial out-degree distributions of the type shown in Fig.~\ref{fig:pkout}.

\pagebreak
\section{Asymptotic equivalence between causal sets in de Sitter spacetime and complex networks in hyperbolic space}\label{sec:theory}

\subsection{De Sitter spacetime}\label{sec:de-sitter}

The $d+1$-dimensional de Sitter spacetime~\cite{GriffithsPodolsky2009} is
the exact solution of the Einstein equations
\begin{equation}\label{eq:einstein}
G_{\mu\nu}+\Lambda g_{\mu\nu} = 0
\end{equation}
for the empty universe with positive vacuum energy density. In these
equations, $G_{\mu\nu}=R_{\mu\nu}-\frac{1}{2}Rg_{\mu\nu}$ is the Einstein
tensor, $R_{\mu\nu}$ the Ricci tensor, $R$ the scalar curvature,
$g_{\mu\nu}$ the metric tensor, $\Lambda$ the cosmological constant, and all
the notations are in the natural units with the speed of light $c=1$.
This spacetime can be represented as the one-sheeted $d+1$-dimensional hyperboloid of constant positive scalar $R$ and Gaussian $K$ curvatures~\cite{Postnikov-book}
\begin{equation}
-z_0^2+z_1^2+\ldots+z_{d+1}^2=a^2=\frac{d(d-1)}{2\Lambda}=\frac{d(d+1)}{R}=\frac{1}{K}
\end{equation}
borrowing its metric from the $d+2$-dimensional ambient Minkowski space with metric
\begin{equation}\label{eq:mikowski-metric}
ds^2 = -dz_0^2+dz_1^2+\ldots+dz_{d+1}^2.
\end{equation}
This hyperboloid in the $d=1$ case is visualized as the outer hyperboloid in Fig.~\ref{fig:2}. As a side note, the curvature of the same hyperboloid in the Euclidean metric is everywhere negative but not constant. For $d=3$, the Hubble constant~$H$, vacuum energy density~$\rho_\Lambda$, cosmological constant~$\Lambda$, and the hyperboloid pseudoradius~$a$, scalar curvature~$R$, and Gaussian curvature~$K$ are all related by
\begin{equation}
H^2 = \frac{8}{3}\pi G\rho_\Lambda = \frac{\Lambda}{3} = \frac{1}{a^2} = \frac{R}{12}=K,
\end{equation}
where $G$ is the gravitational constant.

De Sitter spacetime admits different natural coordinate systems with negative, zero, or positive {\em spatial\/} curvatures, which are not to be confused with the positive curvature of the whole {\em spacetime}. Here we use the standard coordinate system $(t,\theta_1,\ldots,\theta_d)$ with positive spatial curvature that covers the whole spacetime:
\begin{eqnarray}
z_0 &=& a \sinh\frac{t}{a},\label{eq:ds-coord-first}\\
z_1 &=& a \cosh\frac{t}{a}\cos\theta_1,\\
&\vdots&\nonumber\\
z_d &=& a \cosh\frac{t}{a}\sin\theta_1\ldots\sin\theta_{d-1}\cos\theta_d,\\
z_{d+1} &=& a \cosh\frac{t}{a}\sin\theta_1\ldots\sin\theta_{d-1}\sin\theta_d,\label{eq:ds-coord-last}
\end{eqnarray}
where $t\in\R$ is the cosmological time of the universe, and $\theta_1,\ldots,\theta_{d-1}\in[0,\pi]$ and $\theta_d\in[0,2\pi]$ are the standard angular coordinates on the unit $d$-dimensional sphere $\S^d$. The time $t$ at spacetime point $P$ is also the Minkowski length of the arc connecting $P$ to the corresponding point at time $t=0$ belonging to the $z_0=0$ slice of the hyperboloid, see Fig.~\ref{fig:2}. In these coordinates the metric takes the Friedmann–-Lema\^{\i}tre–-Robertson–-Walker (FLRW) form for an exponentially expanding homogeneous and isotropic universe with positive spatial curvature:
\begin{eqnarray}
ds^2 &=& -dt^2 + a^2\cosh^2\frac{t}{a}\left(d\theta^2+\sin^2\theta\,d\Omega_{d-1}^2\right) = -dt^2 + a^2\cosh^2\frac{t}{a}\,d\Omega_d^2,\text{ where}\\
d\Omega_d^2 &=& d\theta_1^2 + \sin^2\theta_1d\theta_2^2+\ldots+\sin^2\theta_1\ldots\sin^2\theta_{d-1}d\theta_d^2
\end{eqnarray}
is the metric on $\S^d$. That is, at each time $t$, the universe is a sphere of exponentially growing radius $a\cosh(t/a)$ and volume
\begin{equation}
v = \sigma_d\left(a\cosh\frac{t}{a}\right)^d, \quad\text{where $\sigma_d = \frac{2\pi^\frac{d+1}{2}}{\Gamma\left(\frac{d+1}{2}\right)}$}
\end{equation}
is the volume of $\S^d$. Figure~\ref{fig:2} visualizes this foliation for $d=1$, in which case these time-slice spheres are circles, and their volume is the circle's circumference.

\begin{figure}
\centerline{\includegraphics[width=3in]{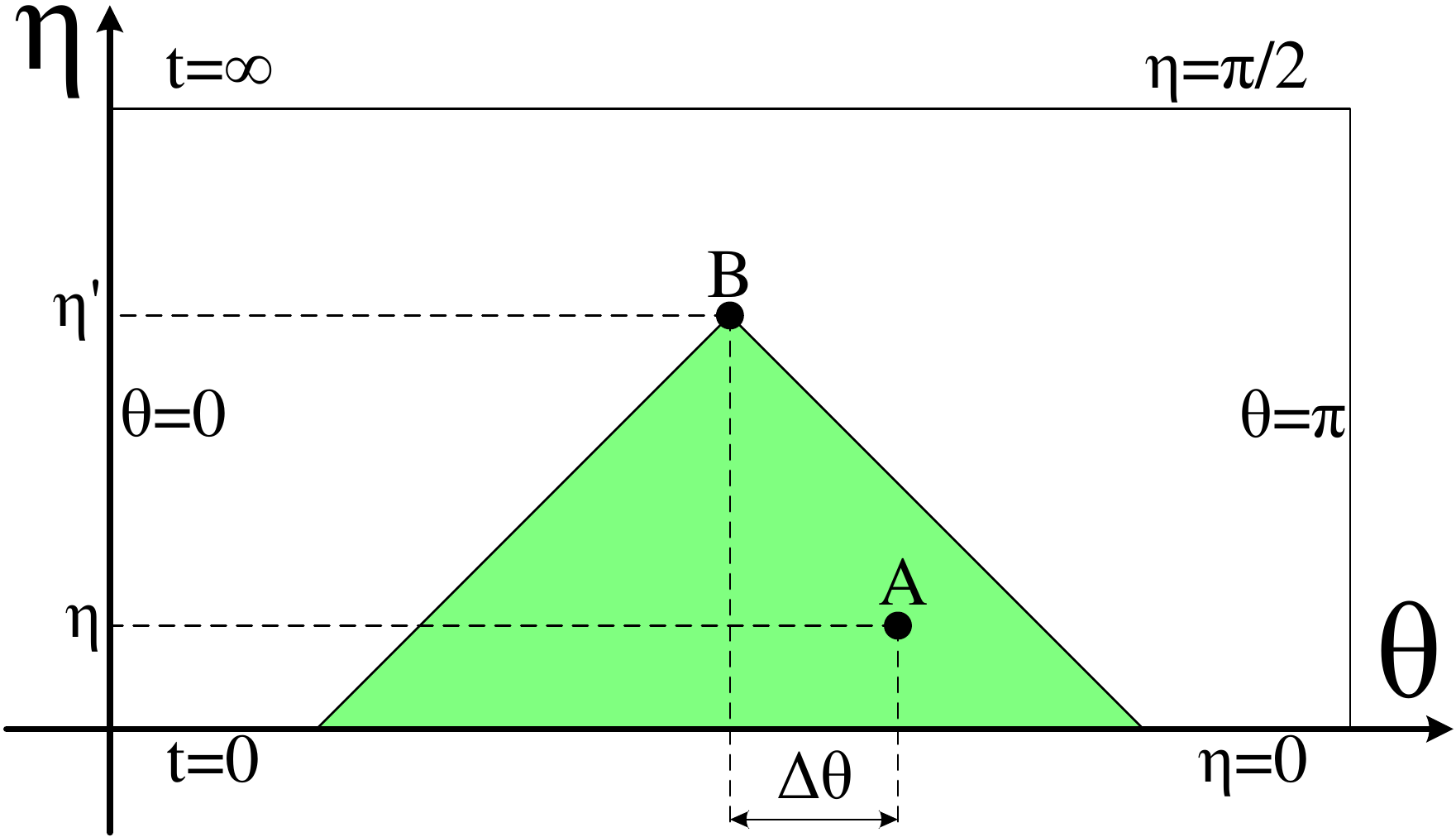}}
\caption{Causal structure of the $d+1$-dimensional de Sitter spacetime. $d-1$
  dimensions are suppressed, so that each
  point represents a $d-1$-sphere.
$\Delta\theta$ is the distance between $A$ and $B$ on the $d$-dimensional unit sphere $\S^d$.}
\label{fig:dsd}
\end{figure}

To study the causal structure of de Sitter spacetime, it is convenient to introduce conformal time $\eta\in(-\frac{\pi}{2},\frac{\pi}{2})$ via
\begin{equation}
\sec\eta=\cosh\frac{t}{a}.
\end{equation}
These conformal-time coordinates are convenient because the metric becomes
\begin{equation}
ds^2 = a^2\sec^2\eta\left(-d\eta^2+d\theta^2+\sin^2\theta\,d\Omega_{d-1}^2\right),
\end{equation}
so that the light cone boundaries defined by $\Delta s = 0$ are straight lines at $45^\circ$ with the $(\eta,\theta)$ axes, see Fig.~\ref{fig:dsd}. Therefore, in this figure, point $A$ at time $\eta\in[0,\eta')$ lies in the past light cone of point $B$ at time $\eta'\in(0,\pi/2)$ if the angular distance $\Delta\theta$ between $A$ and $B$ on $\S^d$ is less than the conformal time difference $\eta'-\eta$ between them:
\begin{eqnarray}
\Delta\theta&<&\eta'-\eta = \arcsec\cosh\frac{t'}{a} - \arcsec\cosh\frac{t}{a},\quad\text{or approximately}\label{eq:past-cone-ineq-exact}\\
\Delta\theta&<&2\left(e^{-\frac{t}{a}}-e^{-\frac{t'}{a}}\right)\approx2e^{-\frac{t}{a}},\label{eq:past-cone-ineq-approx}
\end{eqnarray}
where the last approximation holds for $t' \gg t \gg 1$, and where we have used the approximation $\eta=\arcsec\cosh(t/a)\approx\pi/2-2e^{-t/a}$, which is valid for large $t$.

The volume form on de Sitter spacetime in the conformal and cosmological time coordinates is given by
\begin{eqnarray}
dV &=& \left(a\sec\eta\right)^{d+1}d\eta\,d\Phi_d = \left(a\cosh\frac{t}{a}\right)^ddt\,d\Phi_d \approx \left(\frac{a}{2}\right)^de^{\frac{d}{a}t}dt\,d\Phi_d,\quad\text{where}\label{eq:de-sitter-volume-form}\\
d\Phi_d &=& \sin^{d-1}\theta_1\sin^{d-2}\theta_2\ldots\sin\theta_{d-1}d\theta_1\,d\theta_2\ldots d\theta_d
\end{eqnarray}
is the volume form on $\S^d$, i.e.\ $\int d\Phi_d = \sigma_d$.

\subsection{Hyperbolic space}

The hyperboloid model of the $d+1$-dimensional hyperbolic space~\cite{Bonahon09-book} is represented
by one sheet of the two-sheeted $d+1$-dimensional hyperboloid of constant negative scalar $R$ and Gaussian $K$ curvatures
\begin{equation}
-z_0^2+z_1^2+\ldots+z_{d+1}^2=-b^2=\frac{d(d+1)}{R}=\frac{1}{K}
\end{equation}
borrowing its metric from the $d+2$-dimensional ambient Minkowski space with metric~(\ref{eq:mikowski-metric}).
This hyperboloid in the $d=1$ case is visualized as the inner hyperboloid in Fig.~\ref{fig:2}. As a side note, the curvature of the same hyperboloid in the Euclidean metric is everywhere positive but not constant. The standard coordinate system $(r,\theta_1,\ldots,\theta_d)$ on this hyperboloid is given by
\begin{eqnarray}
z_0 &=& b \cosh\frac{r}{b},\label{eq:h-coord-first}\\
z_1 &=& b \sinh\frac{r}{b}\cos\theta_1,\\
&\vdots&\nonumber\\
z_d &=& b \sinh\frac{r}{b}\sin\theta_1\ldots\sin\theta_{d-1}\cos\theta_d,\\
z_{d+1} &=& b \sinh\frac{r}{b}\sin\theta_1\ldots\sin\theta_{d-1}\sin\theta_d,\label{eq:h-coord-last}
\end{eqnarray}
where $r\in\R_+$ is the radial coordinate, and
$\theta_1,\ldots,\theta_{d-1}\in[0,\pi]$ and $\theta_d\in[0,2\pi]$ are the
standard angular coordinates on the unit $d$-dimensional sphere $\S^d$. The
radial coordinate $r$ of point $P$ on the hyperboloid is the Minkowski
length of the arc connecting $P$ to the hyperboloid vertex, which is the bottom of the inner hyperboloid in Fig.~\ref{fig:2}.
As a side note, the Euclidean length $r_E$ of the same arc is given by the incomplete elliptic integral of the second kind $r_E=-iE(ir,2)$.

\subsection{Hyperbolic model of complex networks}\label{sec:hyperbolic-model}

In the hyperbolic model of complex networks~\cite{PaBoKr11}, networks grow over the $d+1$ dimensional hyperbolic space of Gaussian curvature $K=-1/b^2$ according to the following rule in the simplest case. New nodes $n$ are born one at a time, $n=1,2,3,\ldots$, so that $n$ can be called a network time. Each new node is located at a random position on $\S^d$. That is, the angular coordinates $(\theta_1,\ldots,\theta_d)$ for new nodes are drawn from the uniform distribution on $\S^d$. The radial coordinate of the new node is
\begin{equation}\label{eq:r.vs.n}
r=2\frac{b}{d}\ln\frac{n}{\nu},
\end{equation}
where $\nu$ is a parameter controlling the average degree in the network. Upon its birth, each new node connects to all the nodes lying within hyperbolic distance $r$ from itself. In other words, the connectivity perimeter of new node $n$ at time $n$ is the hyperbolic ball of radius $r$ centered at node $n$. The hyperbolic distance $x$ between two points with radial coordinates $r$ and $r'$ located at angular distance $\Delta\theta$ is given by the hyperbolic law of cosines~\cite{Bonahon09-book}:
\begin{equation}
x = b\,\arccosh\left(\cosh\frac{r}{b}\cosh\frac{r'}{b}-\sinh\frac{r}{b}\sinh\frac{r'}{b}\cos\Delta\theta\right)\approx r+r'+2b\ln\frac{\Delta\theta}{2}.
\end{equation}
Therefore new node $n'$ connects to existing nodes $n<n'$ whose coordinates satisfy
\begin{eqnarray}
x&<&r',\quad\text{or approximately}\label{eq:ball-ineq-exact}\\
r+2b\ln\frac{\Delta\theta}{2}&<&0\label{eq:ball-ineq-approx}.
\end{eqnarray}

This construction yields growing networks whose distribution $P(k)$ of node degrees $k$ is a power law, $P(k)\sim k^{-\gamma}$, with $\gamma=2$. Indeed, according to Eq.~(\ref{eq:r.vs.n}), the radial density of nodes at any given time scales with $r$ as $\rho(r)\sim e^{\alpha r}$, where $\alpha=d/(2b)$. One can also calculate, see~\cite{KrPa10}, the average degree of nodes at radial coordinate $r$, which is $\bar{k}(r)\sim e^{-\beta r}$, where $\beta=\alpha$. The probability that a node at $r$ has degree $k$ is given by the Poisson distribution with the mean equal to $\bar{k}(r)$. Taken altogether, these observations prove that $\gamma=\alpha/\beta+1=2$. The networks in the model also have strongest possible clustering, i.e.\ the largest possible number of triangular subgraphs, for graphs with this degree distribution, and their average degree is
\begin{equation}\label{eq:average-degree}
\bar{k}\approx2^{d+1}\frac{\upsilon_d}{\sigma_d}\nu\ln n,\quad\text{where $\upsilon_d = \frac{\sigma_{d-1}}{d}$}
\end{equation}
is the volume of the unit $d$-dimensional ball. The model and its extensions describe the large-scale structure and growth dynamics of different real networks, e.g.\ the Internet, metabolic networks, and social networks, with a remarkable accuracy~\cite{PaBoKr11}.

\subsection{Duality between de Sitter causal sets and complex networks}

To demonstrate the asymptotic
equivalence between growing
networks from the previous section and causal sets growing in de Sitter
spacetime, we:
\begin{enumerate}
\item find a mapping of points in the hyperbolic space $\H^{d+1}$ to de Sitter spacetime $d\S^{d+1}$ such that:
\item the hyperbolic ball of new node $n\in\H^{d+1}$ is asymptotically identical to its past light cone in $d\S^{d+1}$ upon the mapping, and
\item the distribution of nodes after mapping is uniform in $d\S^{d+1}$.
\end{enumerate}

\subsubsection{Mapping}

A mapping that, as we prove below, satisfies the properties above is remarkably simple:
\begin{eqnarray}
t&=&r,\label{eq:t=r}\\
a&=&2b.
\end{eqnarray}
That is, we identify radial coordinate $r$ in $\H^{d+1}$ with time $t$ in $d\S^{d+1}$, keeping all the angular coordinates the same---see Fig.~\ref{fig:2} for illustration.

\subsubsection{Hyperbolic balls versus past light cones}

The proof that the hyperbolic balls of new node connections map asymptotically to past light cones is trivial: inequalities~(\ref{eq:ball-ineq-approx}) and~(\ref{eq:past-cone-ineq-approx}) are identical with the mapping above.
Figure~\ref{fig:2}(d-f) visualizes the approximation accuracy at different times.

\subsubsection{Uniform node density}

The proof that the node density after mapping is uniform in $d\S^{d+1}$ is trivial as well. Indeed, according to Eq.~(\ref{eq:r.vs.n}) with $2b=a$, network time $n$ is related to cosmological time $t$ via
\begin{eqnarray}
n &=& \nu e^{\frac{d}{a}t},\quad\text{so that}\label{eq:network-time}\\
dn &=& \nu\frac{d}{a}e^{\frac{d}{a}t}\,dt.
\end{eqnarray}
Using the last equation, we rewrite the de Sitter volume element in Eq.~(\ref{eq:de-sitter-volume-form}) as
\begin{equation}
dV = \frac{2b^{d+1}}{\nu d}dn\,d\Phi_d.
\end{equation}
By construction in Section~\ref{sec:hyperbolic-model}, the node density on $\S^d$ is uniform and equal to~$1$. Therefore the number of nodes $dN$ in element $dn\,d\Phi_d$ is
\begin{equation}
dN = \frac{1}{\sigma_d}dn\,d\Phi_d.
\end{equation}
Combining the last two equations, we obtain
\begin{eqnarray}
dN&=&\delta\,dV,\quad\text{where}\\
\delta &=& \frac{\nu d}{2b^{d+1}\sigma_d},\label{eq:delta}
\end{eqnarray}
meaning that nodes are distributed uniformly in $d\S^{d+1}$ with constant density $\delta$, thus completing the proof.

A causal set growing in de Sitter spacetime with node density $\delta$ in Eq.~(\ref{eq:delta}) is thus asymptotically equivalent to a growing complex network in Section~\ref{sec:hyperbolic-model} with average degree $\bar{k}$ in Eq.~(\ref{eq:average-degree}). Combining these two equations, we can relate $\delta$ and $\bar{k}$ to each other:
\begin{equation}\label{eq:ka-desitter}
\bar{k}\approx2\upsilon_d\delta a^dt.
\end{equation}
An important consequence of this asymptotic equivalence is that the degree distribution in both cases is the same power law with exponent $\gamma=2$.

\subsection{Lorentz invariance}

Here we show that the described duality is Lorentz-invariant. The group of isometries of $d\S^{d+1}$ and $\H^{d+1}$ is $SO(d+1,1)$, i.e.\ the Lorentz group of the ambient Minkowski space $\M^{d+2}$. Any $g\in SO(d+1,1)$ preserves volumes, maps light cones in $d\S^{d+1}$ to light cones, and sends balls in $\H^{d+1}$ to balls of the same radii. The fact that the mapping $\Phi:d\S^{d+1}\mapsto\H^{d+1}$ described in the previous section is invariant under the $SO(d+1)$ subgroup of $SO(d+1,1)$ is immediately obvious. Consider now \emph{any} element $g\in SO(d+1,1)$. This element defines a new coordinate system in $\M^{d+2}$, $\tilde{z}_i=\sum_{j=0}^{d+1}g_{ij}z_j$, where $g_{ij}$ is the matrix representation of $g$, e.g.\ if $g$ is the Lorentz boost with rapidity $\phi$ in the $z_1$ direction, then $g_{00}=g_{11}=\cosh\phi$, $g_{12}=g_{21}=-\sinh\phi$, and $g_{kk}=1$, $g_{kl}=0$ for $k,l>1$. These new $\tilde{z}$-coordinates define new coordinates $(\tilde{t},\tilde{\theta}_1,\ldots,\tilde{\theta}_d)$ and $(\tilde{r},\tilde{\theta}_1,\ldots,\tilde{\theta}_d)$ on $d\S^{d+1}$ and $\H^{d+1}$ via the same Eqs.~(\ref{eq:ds-coord-first}-\ref{eq:ds-coord-last},\ref{eq:h-coord-first}-\ref{eq:h-coord-last}). The same~$g$, considered as an active transformation, sends the patch between $t=0$ and $t=t'$ in $d\S^{d+1}$ in Fig.~\ref{fig:2} to \emph{another} patch. As a side note, since $SO(d+1,1)$ is not compact, there exists no compact subset of $d\S^{d+1}$ (or $\H^{d+1}$) invariant under the action of $SO(d+1,1)$. Yet one can check that $g$ sends the patch between $t=0$ and $t=t'$ to the patch between $\tilde{t}=0$ and $\tilde{t}=t'$, the past light cone of point $P$ at $t=t'$---to the past light cone of $P$'s image $\tilde{P}$ at $\tilde{t}=t'$, the ball of radius $r'$ centered at $r=0$ in $\H^{d+1}$---to the ball of the same radius centered at $\tilde{r}=0$, and so on. Therefore, mapping $\tilde{\Phi}:d\S^{d+1}\mapsto\H^{d+1}$ after the transformation induced by $g$ is exactly the same as in Eq.~(\ref{eq:t=r}), i.e.\ $\tilde{t}=\tilde{r}$, and under the action of $SO(d+1,1)$ this mapping transforms as $\tilde{\Phi}=g\circ\Phi\circ g^{-1}$.

\pagebreak
\section{The universe as a causal set}\label{sec:universe}

De Sitter spacetime is the spacetime of a universe with positive vacuum (dark) energy density and no matter or radiation. Since the real universe does contain matter, its spacetime deviates from the pure de Sitter spacetime. At early times, matter dominates, leading to the Big Bang singularity at $t=0$ that de Sitter spacetime lacks. At later times, the matter density decreases, while the dark energy density stays constant, so that it starts dominating, and the universe becomes asymptotically de Sitter. The universe today is at the crossover between the matter-dominated and dark-energy-dominated eras, since the matter and dark energy densities $\rho_M$ and $\rho_\Lambda$ are of the same order of magnitude, leading to rescaled cosmological time $\tau=t/a\sim1$---the so-called ``{\it why now?}'' puzzle in cosmology~\cite{GaLi99,So07,BaSh11,HaSh12}.

To quantify these deviations from pure de Sitter spacetime, and their effect on the structure of the causal set of the real universe, we calculate its degree distribution in this section. This task is quite challenging, and in what follows we first provide the exact analytic expression for the degree distribution, and then derive its approximations based on the measured properties of the universe. These approximations turn out to be remarkably accurate because according to the current measurements, the universe is almost flat.

We emphasize that the approximations in this section are based on the exact solution of the Einstein equations for a flat universe containing only matter and constant positive vacuum energy, i.e.\ constant cosmological constant~$\Lambda$, and that we assume that the universe is governed by this solution at all times. This assumption is a simplification of reality for a number of reasons. For example, there are a plenty of cosmological scenarios, such as eternal inflation~\cite{Kleban2011}, in which the ultimate fate of the universe deviates from the asymptotic solution that we consider. There are also a variety of other models with non-constant $\Lambda$, yet the standard $\Lambda$CDM model with constant $\Lambda$ is a baseline describing accurately many observed properties of the real universe~\cite{Komatsu11}, partly justifying our simplifying assumption. We also note that by relying on the exact solution for a universe containing only matter and $\Lambda$, we effectively neglect the earliest stages of universe evolution such as the radiation-dominated era or inflationary epoch. There is no consensus on how exactly the universe evolved at those earliest times.  Since the radiation-dominated era ended soon after the Big Bang~\cite{Woodard2009}, these details are unlikely to have a profound effect on the universe causet's structure at much later times.

\subsection{Exact expression for the degree distribution}

The metric in a homogeneous and isotropic universe takes the FLRW form:
\begin{eqnarray}
ds^2 &=& -dt^2 + R^2(t)\left\{d\chi^2+\frac{1}{K}\sin^2\left[\sqrt{K}\chi\right]\left(\sin^2\theta\,d\theta^2+d\phi^2\right)\right\},\text{ where}\label{eq:FLRWmetric}\\
\frac{1}{K}\sin^2\sqrt{K}\chi&=&
\begin{cases}
\sin^2\chi&\text{if $K=1$ (closed universe with positive spatial curvature),}\\
\chi^2&\text{if $K=0$ (flat universe with zero spatial curvature),}\\
\sinh^2\chi&\text{if $K=-1$ (open universe with negative spatial curvature).}
\end{cases}
\end{eqnarray}
In the above expression, coordinates $\phi\in[0,2\pi]$ and $\theta\in[0,\pi]$ are the standard angular coordinates on $\S^2$, while $\chi\in[0,\pi]$ if $K=1$, or $\chi\in[0,\infty]$ if $K=0,-1$, is the radial coordinate in a spherical, flat, or hyperbolic space. Finally, $t\in[0,\infty]$ is the cosmological time, and $R(t)$ is the scale factor, finding an appropriate approximation to which in the real universe is an important part of our approximations in subsequent sections. In this and the next sections, however, all the expressions are valid for {\em any\/} scale factor.

As with pure de Sitter, it is convenient to introduce conformal time $\eta$, related to cosmological time $t$ via
\begin{equation}
\eta = \int_0^t\frac{dt'}{R(t')}.
\end{equation}
In conformal time coordinates, the metric and volume form become
\begin{eqnarray}
ds^2 &=& R^2(\eta)\left\{-d\eta^2+d\chi^2+\frac{1}{K}\sin^2\left[\sqrt{K}\chi\right]\left(\sin^2\theta\,d\theta^2+d\phi^2\right)\right\},\\
dV &=& \frac{1}{K}R^4(\eta)\sin^2\left[\sqrt{K}\chi\right]\sin\theta\,d\eta\,d\chi\,d\theta\,d\phi.
\end{eqnarray}

As with pure de Sitter, we orient edges in the universe's causet from future to the past. Therefore if $\eta$ is the current conformal time,
then the average in- and out-degrees $\bar{k}_{i,o}(\eta'|\eta)$ of nodes born at conformal time $\eta'<\eta$ are simply proportional to the volumes of their future and past light cones $V_{f,p}(\eta'|\eta)$:
\begin{eqnarray}
\bar{k}_i(\eta'|\eta) &=& \delta V_f(\eta'|\eta),\\
\bar{k}_o(\eta'|\eta) &=& \delta V_p(\eta'|\eta),
\end{eqnarray}
where the coefficient of proportionality is the Planck-scale node density in the spacetime, which
we take to be the inverse of the Planck $4$-volume:
\begin{eqnarray}\label{eq:planck_time}
\delta &=& \frac{1}{t_P^4} = 1.184 \times 10^{173} \text{ s$^{-4}$, where}\\
t_P &=& 5.391 \times 10^{-44}\text{ s}
\end{eqnarray}
is the Planck time.

Thanks to conformal time coordinates, the expressions for volumes $V_{f,p}(\eta'|\eta)$ are easy to write down:
\begin{eqnarray}
V_f(\eta'|\eta) &=& \frac{1}{K}\int_{\eta'}^{\eta} d\eta''\,R^4(\eta'')\int_0^{\eta''-\eta'}d\chi\,\sin^2\left[\sqrt{K}\chi\right]
\int_0^\pi d\theta\,\sin\theta\int_0^{2\pi}d\phi,\\
V_p(\eta'|\eta) &=& \frac{1}{K}\int_0^{\eta'} d\eta''\,R^4(\eta'')\int_0^{\eta'-\eta''}d\chi\,\sin^2\left[\sqrt{K}\chi\right]
\int_0^\pi d\theta\,\sin\theta\int_0^{2\pi}d\phi.
\end{eqnarray}
Computing the three inner integrals, we obtain
\begin{eqnarray}
\bar{k}_i(\eta'|\eta) &=& \delta\frac{\pi}{K}\int_{\eta'}^{\eta}\left\{2(\eta''-\eta')-\frac{1}{\sqrt{K}}\sin\left[2\sqrt{K}(\eta''-\eta')\right]\right\}R^4(\eta'')\,d\eta'',\label{eq:ki(eta)}\\
\bar{k}_o(\eta'|\eta) &=& \delta\frac{\pi}{K}\int_{0}^{\eta'}\left\{2(\eta'-\eta'')-\frac{1}{\sqrt{K}}\sin\left[2\sqrt{K}(\eta'-\eta'')\right]\right\}R^4(\eta'')\,d\eta''.\label{eq:ko(eta)}
\end{eqnarray}
As shown in~\cite{BoHe09}, Lorentz invariance implies that nodes/events of the causet are distributed in spacetime according to a Poisson point process. Therefore, as explained in~\cite{BoPa03}, to find the in- or out-degree distributions $P(k,\eta)$ at time~$\eta$, we have to average the Poisson distribution
\begin{equation}
p(k|\eta',\eta) = \frac{1}{k!}\left[\bar{k}(\eta'|\eta)\right]^{k} e^{-\bar{k}(\eta'|\eta)},
\end{equation}
which is the probability that a node born at time $\eta'$ has degree $k$, with the density of nodes born at time $\eta'$
\begin{eqnarray}
\rho(\eta'|\eta) &=& \frac{R^4(\eta')}{{\cal{N}}(\eta)},\text{ where}\\
{\cal{N}}(\eta)  &=& \int_0^{\eta} R^4(\eta') d\eta'
\end{eqnarray}
is the time-dependent normalization factor. The result is
\begin{equation}
P(k,\eta) = \int_0^\eta p(k|\eta',\eta)\rho(\eta'|\eta)\,d\eta'
=\frac{1}{\cal{N}(\eta)} \frac{1}{k!} \int_0^{\eta} \left[\bar{k}(\eta'|\eta)\right]^{k} e^{-\bar{k}(\eta'|\eta)} R^4(\eta') \, d\eta'.
\label{eq:degreedistribution}
\end{equation}
The above expressions are valid for both in-degree (set $k \equiv k_i$ and $\bar{k}(\eta'|\eta) \equiv \bar{k}_i(\eta'|\eta)$) and out-degree ($k \equiv k_o$, $\bar{k}(\eta'|\eta) \equiv \bar{k}_o(\eta'|\eta)$) distributions. In what follows we focus on the in-degree distribution.

\subsection{First approximation using the Laplace method}\label{sec:laplace}

Equations~(\ref{eq:degreedistribution},\ref{eq:ki(eta)}) give the exact solution for the in-degree distribution with an arbitrary scale factor $R(\eta)$, but it is difficult to extract any useful information from these expressions, even if we know the exact form of $R(\eta)$. The first step to get a better insight into the causet properties is to rewrite $\bar{k}_i(\eta'|\eta)$ as
\begin{equation}
\bar{k}_i(\eta'|\eta)=\bar{k}_i(0|\eta) F(\eta'|\eta),\quad\text{where $F(\eta'|\eta)=\frac{\bar{k}_i(\eta'|\eta)}{\bar{k}_i(0|\eta)}$.}
\end{equation}
The in-degree of the oldest node $\bar{k}_i(0|\eta)$ is an astronomically large number, both because the node is old, so that its future light cone comprises a macroscopic portion of the $4$-volume of the whole universe, and because the degree is proportional to huge $\delta$. However, function $F(\eta'|\eta)$ is a monotonically decreasing function whose values lie in the interval $[0,1]$. We can thus use the Laplace method to approximate the integral in Eq.~(\ref{eq:degreedistribution}) in the limit $\delta\gg1$. Introducing $\eta^*(k_i)$, which is the solution of the transcendent equation
\begin{equation}
k_i=\bar{k}_i(\eta^*|\eta),
\end{equation}
and function
\begin{equation}\label{eq:Phi}
\Phi(k_i,\eta)=\left|\frac{\partial\bar{k}_i(\eta'|\eta)}{\partial\eta'}\right|_{\eta'=\eta^*(k_i)},
\end{equation}
the result of this Laplace approximation reads:
\begin{equation}
P(k_i,\eta)=
\begin{cases}
\displaystyle{
\left(\frac{3}{\cal{\pi \delta}} \right)^\frac{1}{4} \Gamma\left(\frac{5}{4}\right) \frac{R^3(\eta)}{{\cal{N}}(\eta)}
}
&\text{if $k_i=0$,}\\[0.7cm]
\displaystyle{
\frac{1}{\cal{N}(\eta)} \frac{\sqrt{2 \pi k_i} k_i^{k_i} e^{-k_i}}{k_i!} \frac{R^4[\eta^*(k_i)]}{\Phi(k_i,\eta)}
\approx\frac{1}{\cal{N}(\eta)}\frac{R^4[\eta^*(k_i)]}{\Phi(k_i,\eta)}
}
&\text{if $1 \le k_i \le \bar{k}_i(0|\eta)$,}
\end{cases}
\label{eq:P(k)approx}
\end{equation}
where we have also used Stirling's approximation $k!\approx\sqrt{2\pi k}(k/e)^k$. We see that the shape of the degree distribution is almost fully determined by $\eta^*(k_i)$, and that  the distribution is a fast decaying function for degrees above $\bar{k}_i(0|\eta)$.

The expression for the in-degree distribution in Eq.~(\ref{eq:P(k)approx}) is now more tractable and gets ready to accept the scale factor $R(\eta)$ of the universe. Unfortunately, the exact expressions for the scale factor of a closed or open universe with matter and dark energy, although known~\cite{Edwards72}, resist analytic treatment, and so does the integral for the average degree $\bar{k}_i(\eta'|\eta)$ in Eq.~(\ref{eq:ki(eta)}) that we are to use in Eq.~(\ref{eq:P(k)approx}). Therefore we next develop a series of approximations to the scale factor and average degree, based on the measured properties of the universe.

\subsection{Measured properties of the universe}\label{sec:measurements}

The current measurements of the universe~\cite{Komatsu11} that are relevant to us here include:
\begin{eqnarray}
\Omega_\Lambda &=& \frac{8\pi G}{3H_0^2}\rho_{\Lambda} \in [0.709,0.741],\text{ (dark energy density)}\label{eq:omega_lambda}\\
\Omega_M &=& \frac{8\pi G}{3H_0^2}\rho_{M} \in [0.2582,0.2914],\text{ (matter density)}\label{eq:omega_m}\\
\Omega_K &=& -\frac{K}{R_0^2H_0^2} \in [-0.0133,0.0084],\text{ (curvature density),\quad where}\label{eq:omega_k}\\
H_0 &=& \frac{\dot{R}_0}{R_0} \in [68.8,71.6]\frac{\text{km}}{\text{s}\cdot\text{Mpc}} = [2.23,2.32]\times 10^{-18}\text{ s$^{-1}$ is the Hubble constant,}\\
R_0 &=& R(\T) \text{ is the scale factor at present time $\T$, i.e.\ the age of the universe,}\\
\T &\in& [13.65,13.87]\text{ Gyr} = [4.308,4.377]\times10^{17}\text{ s, so that}\label{eq:t0-measured}\\
\Lambda &=& 3H_0^2\Omega_\Lambda\in[1.06,1.20]\times10^{-35}\text{ s$^{-2}$, is the cosmological constant, and}\\
a &=& \sqrt{\frac{3}{\Lambda}}=\frac{1}{\sqrt{\Omega_\Lambda}H_0}\in[5.00,5.32]\times10^{17}\text{ s, is the de Sitter pseudoradius.}\label{eq:a(lamda)}
\end{eqnarray}
The range of values of $\Omega_\Lambda$, $\Omega_M$, $H_0$, and $\T$ are given by the $95\%$ confidence bounds in the universe measurements taken from the last column of Table~1 in~\cite{Komatsu11}, while the values of $\Omega_K$ come from Table~2 there: third row, last column.
The matter density $\Omega_M$ is the sum of two contributions: the observable (baryon) matter density ($\Omega_b\in[0.0442,0.0474]$), and dark matter density ($\Omega_c\in[0.214,0.244]$). The radiation density is negligible. The universe thus consists of dark energy ($\approx73\%$), dark matter ($\approx23\%$), and observable matter ($\approx4\%$). As a side note, the sum of all $\Omega$'s $\sum\Omega\in[0.95,1.04]$ as expected, since $\sum\Omega=1$ is the Einstein/Friedmann equation for a FLRW universe, which we recall in the next section. Much more important for us here is that the universe is almost flat, $\Omega_K\approx0$.

\subsection{Approximations to the scale factor and average degree}\label{sec:approximations}

In this section we utilize the approximate flatness of the universe to derive and quantify approximations to the scale factor of the universe and average degree in Eq.~(\ref{eq:ki(eta)}). The main idea is that since $\Omega_K$ is small, we can use the exact solution for the scale factor in a flat universe ($\Omega_K=0$) with matter and dark energy (positive cosmological constant), which is actually quite simple~\cite{GriffithsPodolsky2009}.

\begin{figure}
\centerline{\includegraphics[width=3.5in]{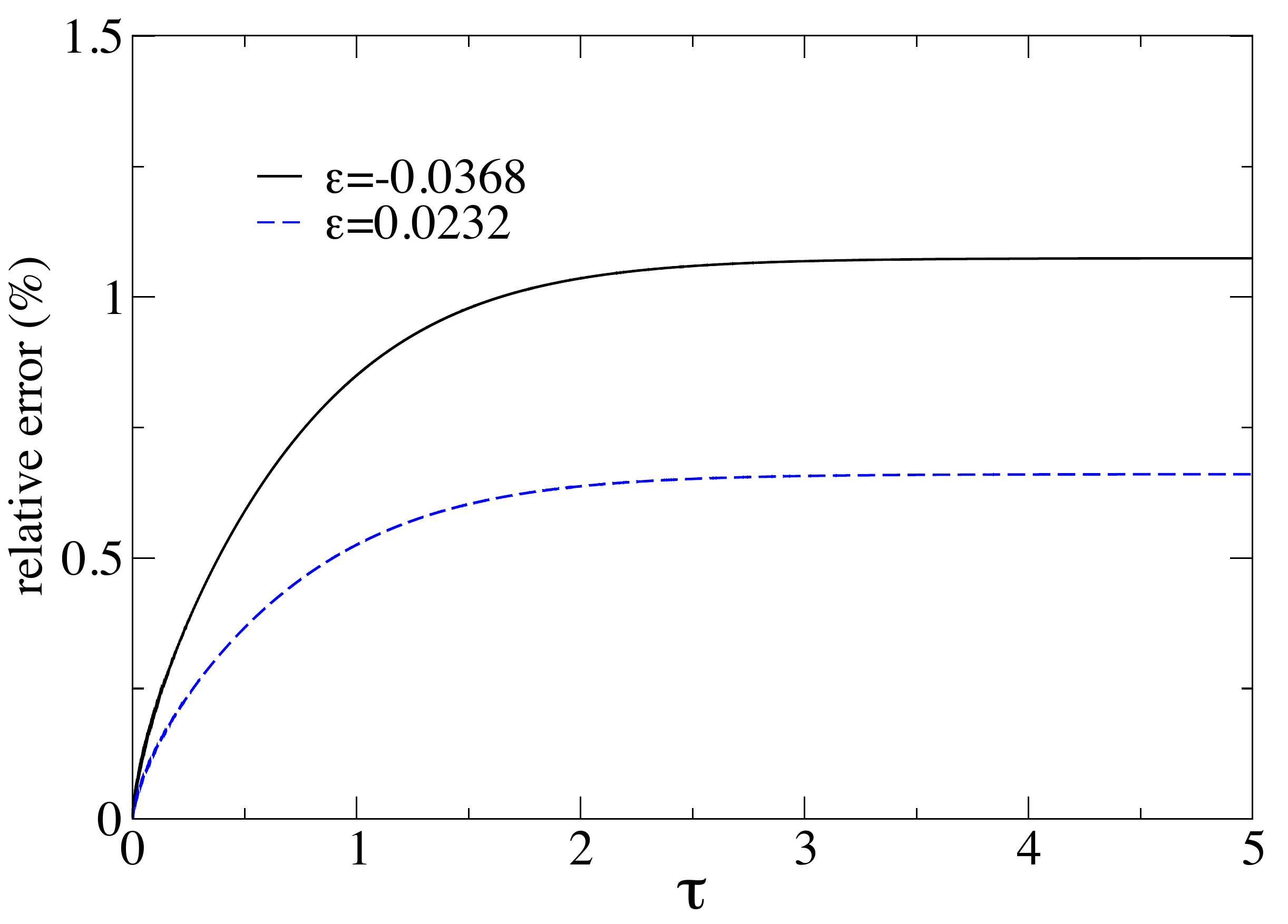}
\includegraphics[width=3.5in]{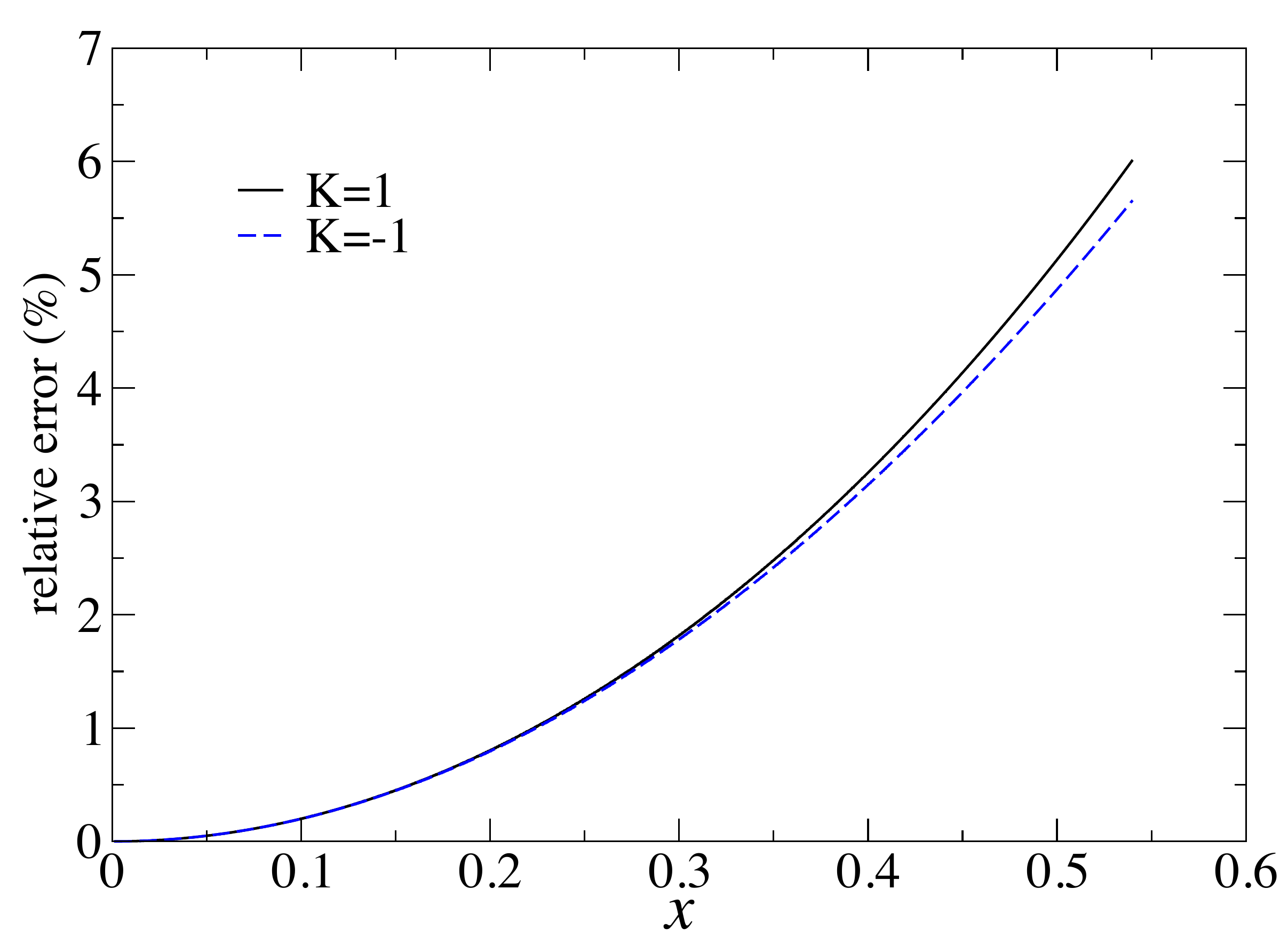}}
\caption{Left: Relative error between functions $r(\tau;\epsilon)$ and $r(\tau;0)$, $|r(\tau;\epsilon)-r(\tau;0)|/r(\tau;\epsilon)$, for the two extremes of the $95\%$ confidence interval of parameter $\epsilon$ in Eq.~(\ref{eq:epsilon}), as a function of rescaled time $\tau$. In the worst case, the relative error is around~$1\%$.
Right: Relative error of approximation in Eq.~(\ref{eq:sin-approximation}) for $x\in[0,\eta_\infty]$.}
\label{fig:accuracy}
\end{figure}

We first recall that the $00$-component of the Einstein equations in the FLRW metric is the first Friedmann equation:
\begin{equation}
\frac{\dot{R}^2+K}{R^2} = \frac{8}{3}\pi G \left[\rho_\Lambda + \left(\frac{R_0}{R}\right)^3\rho_M \right].
\end{equation}
This equation can be rewritten~\cite{Weinberg08-book} in the following form:
\begin{equation}
H_0t = \int_0^{R/R_0}\frac{dx}{x\sqrt{\Omega_\Lambda+\Omega_Kx^{-2}+\Omega_Mx^{-3}}}.\label{eq:scale-factor-integral}
\end{equation}
Assuming non-vanishing positive dark energy and matter densities ($\Omega_\Lambda>0$ and $\Omega_M>0$), using rescaled time
\begin{equation}
\tau\equiv\frac{t}{a},
\end{equation}
and introducing two parameters
\begin{eqnarray}
\alpha &\equiv& R_0 \sqrt[3]{\frac{\Omega_M}{\Omega_{\Lambda}}},\text{ and}\\
\epsilon &\equiv& \frac{\Omega_K}{\sqrt[3]{\Omega_{\Lambda}\Omega_M^{2}}}\in[-0.0368,0.0232],\label{eq:epsilon}
\end{eqnarray}
the integral in Eq.~(\ref{eq:scale-factor-integral}) simplifies to
\begin{equation}
\tau= \int_0^{r}\frac{dx}{x\sqrt{1+\epsilon x^{-2}+x^{-3}}}.
\end{equation}
Solving it for $r \equiv r(\tau,\epsilon)$, we conclude that the solution for the scale factor becomes
\begin{equation}
R(t)=\alpha\,r\left(\tau;\epsilon\right).\label{eq:scale-factor-universe}
\end{equation}
This is definitely not the only way to write down the scale factor in terms of the parameters of the universe. Yet written in this form, the rescaled scale factor $r(\tau;\epsilon)$ is a dimensionless function of its dimensionless arguments, and parameter $\epsilon$ is close to zero, so that $r(\tau;\epsilon) \approx r(\tau;0)$, where $r(\tau,0)$, i.e.\ the rescaled scale factor for a flat universe with positive matter and dark energy densities, is quite simple:
\begin{equation}
r(\tau;0) = \sinh^\frac{2}{3}\frac{3}{2}\tau.\label{eq:sinh23}
\end{equation}
Figure~\ref{fig:accuracy}(left) shows a good agreement between the numerical solution for $r(\tau;\epsilon)$ and the analytical expression for $r(\tau;0)$ in Eq.~(\ref{eq:sinh23}) for the range of the values of $\epsilon$ allowed by measurements in~Eq~(\ref{eq:epsilon}). Therefore in what follows we can use Eq.~(\ref{eq:sinh23}), even if the universe is ``slightly closed'' or ``slightly open.''

To approximate the average degree $\bar{k}_i(\eta'|\eta)$ in Eq.~(\ref{eq:ki(eta)}), we first compute, using the scaling Eq.~(\ref{eq:scale-factor-universe}), the conformal time
\begin{equation}\label{eq:eta}
\eta = \int_0^t\frac{dt'}{R(t')}=\sqrt{|\epsilon|}  \int_0^\tau\frac{d\tau'}{r(\tau';\epsilon)}
= \sqrt{|\epsilon|} \int_0^{r(\tau;\epsilon)} \frac{dx}{x^2 \sqrt{1+\epsilon x^{-2}+x^{-3}}}
\end{equation}
at the future infinity $t=\tau=r(\tau;\epsilon)=\infty$. For the extreme values of $\epsilon$ allowed by measurements, we obtain $\eta_\infty=0.5403$ for the closed universe with $\epsilon=-0.0368$ and $\eta_{\infty}=0.4260$ for the open universe with $\epsilon=0.0232$. If $\epsilon\to0$, then we also have $\eta_\infty\to0$, yet if $\epsilon$ is exactly zero, then $\eta_\infty$ is an undefined constant, since in the flat case, $R_0$ is a free parameter that can be set to an arbitrary value without affecting anything. If the universe is not exactly flat, then the above values of $\eta_\infty$ allowed by measurements are well below $\pi/2$, so that we can safely replace expression $\left[2x-\sin(2\sqrt{K}x)/\sqrt{K}\right]/K$ in Eq.~(\ref{eq:ki(eta)}) by its Taylor expansion around zero,
\begin{equation}\label{eq:sin-approximation}
\frac{1}{K}\left[2x-\frac{1}{\sqrt{K}}\sin(2\sqrt{K}x)\right]\approx\frac{4}{3}x^3.
\end{equation}
Indeed the maximum possible relative error between the left and right hand sides in the last equation at $x=\eta_\infty$ is around $6\%$, see Fig.~\ref{fig:accuracy}(right), whereas such error is zero for an exactly flat universe with $\epsilon=K=0$, since the left and right hand sides are equal in this case. Therefore the approximation to the average degree in Eq.~(\ref{eq:ki(eta)})
\begin{equation}\label{eq:k(eta'|eta)-approx}
\bar{k}_i(\eta'|\eta) = \frac{4}{3}\pi\delta\int_{\eta'}^{\eta}\left(\eta''-\eta'\right)^3R^4(\eta'')\,d\eta''
\end{equation}
is exact for flat universes, and almost exact for slightly closed or open universes with $\epsilon$ within the range of Eq.~(\ref{eq:epsilon}).

\subsection{Scaling relations}\label{sec:scaling}

To proceed to our final approximations to the degree distribution in the next section, we derive some scaling relations that tremendously simplify the calculations, provide important insights into the properties of the degree distribution, and suggest methods to simulate the causal set of the universe.

We begin with the scaling relation for conformal time $\eta$. Let us introduce the rescaled conformal time
\begin{equation}
\zeta \equiv \frac{\alpha}{a}\eta = \frac{\eta}{\sqrt{|\epsilon|}},
\end{equation}
and let notations $y=f_{1,2,\ldots}(x)$ mean ``$y$ is a function of $x$.'' Given the definition of conformal time in Eq.~(\ref{eq:eta}), this rescaled conformal time is a function of rescaled cosmological time $\tau$:
\begin{equation}
\zeta = f_1(\tau) = \int_0^{r(\tau;\epsilon)} \frac{dx}{x^2 \sqrt{1+\epsilon x^{-2}+x^{-3}}}
\xrightarrow[\epsilon\to0]{}2\sqrt{r(\tau;0)}\,\,{}_2F_1\left(\frac{1}{6},\frac{1}{2},\frac{7}{6},-r^3(\tau;0)\right),
\label{eq:zeta}
\end{equation}
where $r(\tau;0)$ is given by Eq.~(\ref{eq:sinh23}), and $_2F_1$ is the hypergeometric function. This scaling means that
\begin{equation}
R(\eta)=\alpha f_2(\zeta),
\end{equation}
which in turn implies that the average in-degree of nodes born at time $\eta'$
\begin{eqnarray}
\bar{k}_i(\eta'|\eta)&=&q f_3\left(\zeta',\zeta\right)=q f_4\left(\tau',\tau\right),\label{eq:ki-scaling}\text{ where}\\
q&\equiv&\delta a^4,\label{eq:Psi}
\end{eqnarray}
while the average in- or out-degree in the causet
\begin{equation}
\bar{k}_i(\eta) = \bar{k}_o(\eta) = \int_0^{\eta_\infty}\bar{k}_i(\eta'|\eta)\rho(\eta'|\eta)\,d\eta' = q f_5\left(\zeta\right)=q f_6\left(\tau\right).
\end{equation}
Plugging scaling Eqs.~(\ref{eq:ki-scaling},\ref{eq:Psi}) into the expression for the in-degree distribution in Eq.~(\ref{eq:P(k)approx}), we conclude that this distribution scales for $k_i \ge 1$ as
\begin{eqnarray}
P(k_i,t) &=& \frac{1}{q} Q\left(\kappa_i,\tau\right),\text{ where}\label{eq:scalingP(k_in)}\\
\kappa_i&\equiv&\frac{k_i}{q},
\end{eqnarray}
and the rescaled in-degree distribution $Q(\kappa_i,\tau)$ is some function of rescaled degree~$\kappa_i$ and time~$\tau$.
This result is important for several reasons:
\begin{enumerate}
\item
It defines a characteristic degree $q=\delta a^4$ and characteristic time $a$ such that the degree distribution depends only on the rescaled dimensionless variables
$\kappa_i=k_i/q$ and $\tau=t/a$. Therefore we are free to set $a=1$ and $\delta=1$, and the remaining task is to find an explicit form of the rescaled distribution $Q(\kappa_i,\tau)$ in Eq.~(\ref{eq:scalingP(k_in)}).
\item
Even though the scale factor in Eq.~(\ref{eq:scale-factor-universe}) does depend on $\alpha$, the in-degree distribution $P(k_i,t)$ does not depend on $\alpha$. This is important from the practical point of view, because we can carry out all the calculations setting $\alpha=1$ as well.
\item
Having set $\alpha=a=\delta=1$, the in-degree distribution explicitly does not depend on any physical parameters of the universe. The causets in flat or open universes are graphs with obviously infinite numbers of nodes from the very beginning of the universe, while the causets of closed universes are always finite, and the number of nodes in them depend on time and on all the three parameters $\alpha$, $a$, and $\delta$. Yet the in-degree distributions in all these causets at a given time are all the same, and in particular, do not depend on the number of nodes at all. This somewhat unexpected result deserves further explanation. We note that parameter $\alpha=R_0\sqrt[3]{\Omega_M/\Omega_{\Lambda}}$ depends on the physical properties of the universe. In particular, if the universe is closed, it defines, via $R_0$ in Eq.~(\ref{eq:omega_k}), the radius of the three-sphere, i.e.\ of the spatial part of the FLRW metric. Since the degree distribution does not depend on $\alpha$, we can freely take the limit $\alpha \rightarrow \infty$ corresponding to a flat space, without affecting the degree distribution. In other words, we are free to choose a closed universe with $K=+1$ in Eq.~(\ref{eq:FLRWmetric}), and this choice can be considered as a degree-distribution-preserving compactification of an infinite flat or hyperbolic space with an infinite number of nodes in it, into a compact spherical space with a finite number of nodes.
\item
The previous point informs us how to simulate the universe causet on a computer. We cannot simulate infinite causets for flat or open universes, but since the degree distribution is the same in closed universes, we are free to simulate the latter. One can check that in a closed universe, the number of nodes in a causet grows with time as
\begin{equation}
N(t)=\delta a \alpha^3 f_7\left(\tau\right).
\end{equation}
We can set $a=1$, and since the in-degree distribution does not depend on the number of nodes $N$, then for any given time $t=a\tau=\tau$ and for any given value of $\delta$, we can generate graphs with any desired size $N$ by choosing the value of $\alpha$ accordingly. The degree distribution obtained from this simulation is then to be re-scaled according to Eq.~(\ref{eq:scalingP(k_in)}) to obtain the rescaled distribution $Q(\kappa_i,\tau)$. Keeping fixed $t=\tau$ and $N$, we can also explore different regions of $Q(\kappa_i,\tau)$ by changing the value of $\delta$.
\end{enumerate}

We have to emphasize that the points above, and the scaling relations in this section, are valid for non-flat universes only when $\eta_{\infty}$ is well below $\pi/2$, so that the approximation in Eq.~(\ref{eq:sin-approximation}) is valid for all $x\in[0,\eta_\infty]$. When this condition does not hold, e.g.\ when $\epsilon$ is large and the universe is far from being flat, then all these scaling relations break down. In particular, we cannot claim that in that case the in-degree distribution would still be the same. The latest measurements of the universe parameters indicate that the scaling laws derived in this section do apply to the real universe, but even if this were not the case, these laws would still be valid for small values of conformal time.

\subsection{Final approximations to the degree distribution in the matter- and dark-energy-dominated eras}

We now have all the material needed to derive the final approximations to the degree distribution. According to Section~\ref{sec:approximations} the exact solution for the scale factor in a flat universe is a good approximation for the scale factor in the real universe within the measurement-allowed range of parameter values, so that we will use this approximation here, i.e.\ we set $\epsilon=0$. Using the scaling results from Section~\ref{sec:scaling}, we also set $a=\alpha=\delta=1$, which is equivalent to working with rescaled cosmological time $\tau$, rescaled conformal time $\zeta$, rescaled scale factor $r(\tau)=\sinh^{2/3}{(3\tau/2)}$, and rescaled in-degree $\kappa_i$. Since everything is rescaled in this section, we omit word ``rescaled'' in front of the variable names.

We consider two eras of the universe evolution. The first era characterizes the causal set of the universe at early times, when matter dominates. The second era deals with the aged universe at large times, when dark energy dominates. In the limit of infinite time (future infinity), we derive the exact solution for the degree distribution in a flat universe.

\subsubsection{Matter-dominated era}

In this era with $\tau\ll1$, the scale factor and conformal time can be approximated by
\begin{eqnarray}
r(\tau)&=&\left(\frac{3}{2}\tau\right)^\frac{2}{3}\\
\zeta&=&2\left(\frac{3}{2}\tau\right)^\frac{1}{3},\text{ so that}\\
r(\zeta)&=&\left(\frac{\zeta}{2}\right)^2.
\end{eqnarray}
Using these expressions in the rescaled version of Eq.~(\ref{eq:k(eta'|eta)-approx}), we obtain the average in-degree at time $\zeta$ of nodes born at time $\zeta'$:
\begin{equation}\label{eq:kappa(zeta'|zeta)-matter}
\bar{\kappa}_i(\zeta'|\zeta)=\frac{\pi}{768} \zeta^{12} \left[ \frac{1}{12}- \frac{3}{11} \frac{\zeta'}{\zeta}+\frac{3}{10} \left(\frac{\zeta'}{\zeta}\right)^2- \frac{1}{9} \left(\frac{\zeta'}{\zeta}\right)^3+\frac{1}{1980} \left(\frac{\zeta'}{\zeta}\right)^{12}\right].
\end{equation}
The maximum average in-degree, i.e.\ the average in-degree of the oldest nodes, is
\begin{equation}
\bar{\kappa}_i(0|\zeta)=\frac{\pi}{9216} \zeta^{12} = 9\pi\tau^4.
\end{equation}
We note that if we reinsert constants $a$ and $\delta$, and cosmological time $t$ using Eq.~(\ref{eq:ki-scaling}), the  average in-degree of the oldest nodes becomes
\begin{equation}
\bar{k}_i(0|t)=9\pi \left( \frac{t}{t_P} \right)^4,
\end{equation}
where $t_P$ is the Planck time, see Eq.~(\ref{eq:planck_time}). As expected, this in-degree does not depend on $a$ and, consequently, on the cosmological constant $\Lambda$.

For degrees well below the maximum degree, $\kappa_i\ll9\pi\tau^4$, the Taylor expansion of Eq.~(\ref{eq:kappa(zeta'|zeta)-matter}) around $\zeta'=\zeta$ yields
\begin{equation}
\bar{\kappa}_i(\zeta'|\zeta) \approx \frac{\pi}{3072}  \zeta^{12} \left( \frac{\zeta'}{\zeta}-1\right)^4.
\end{equation}
Using this expression in the rescaled version of Eq.~(\ref{eq:P(k)approx}), we obtain
\begin{equation}
Q(\kappa_i,\tau)=\frac{3}{4} \left(\frac{3}{\pi}\right)^\frac{1}{4} \frac{\kappa_i^{-3/4}}{\tau}\quad\text{if $\kappa_i \ll 9\pi \tau^4$},
\end{equation}
and, after reinserting all the physical constants,
\begin{equation}
P(k_i,t)=\frac{3}{4} \left(\frac{3}{\pi}\right)^\frac{1}{4} \frac{t_P}{t} k_i^{-3/4}\quad\text{if $1 \leq k_i \ll 9\pi \left(\frac{t}{t_P}\right)^4$},
\end{equation}
with a soft cut-off at $9\pi \left(t/t_P\right)^4$. We thus observe that the degree distribution does not depend on the cosmological constant $\Lambda$, and that it is a power law $P(k_i,\tau)\sim k_i^{-\gamma}$ with exponent $\gamma=3/4$.

\subsubsection{Dark-energy-dominated era}

We now analyze the future fate of the universe at $\tau\gg1$, which is slightly more intricate. To begin, we derive a couple of expressions that we use in simulations and that are valid for any $\tau$. We first recall that, according to Eq.~(\ref{eq:zeta}), conformal time is related to the current value of the scale factor $r$ via
\begin{equation}\label{eq:zeta(r)-energy}
\zeta(r)=2 \sqrt{r}\,\,{}_2F_1\left(\frac{1}{6},\frac{1}{2},\frac{7}{6},-r^3 \right),
\end{equation}
which is a monotonously increasing function of $r$. Therefore we can use scale factor value $r$ as a measure of time instead of $\zeta$. Using this observation in Eq.~(\ref{eq:k(eta'|eta)-approx}), we write the average in-degree at time $r$ of nodes born at time $r'$ as
\begin{equation}
\bar{\kappa}_i(r'|r)=\frac{4\pi}{3} \int_{r'}^r \left[\zeta(x)-\zeta(r') \right]^3  \frac{x^2 dx}{\sqrt{1+x^{-3}}}.
\label{eq:kappa(r)}
\end{equation}
Similarly, function $\Phi(\kappa_i,r)$ in Eq.~(\ref{eq:Phi}) becomes
\begin{equation}
\Phi(\kappa_i,r)=4 \pi \int_{r^*(\kappa_i)}^r \left[\zeta(x)-\zeta(r^*(\kappa_i)) \right]^2  \frac{x^2 dx}{\sqrt{1+x^{-3}}},
\label{eq:phi(r)}
\end{equation}
where $r^*(\kappa_i)$ is the solution of equation $\bar{\kappa}_i(r^*|r)=\kappa_i$.

Assuming now that time is large, we see from Eq.~(\ref{eq:kappa(r)}) that the maximum average in-degree scales at $r\gg1$ as
\begin{eqnarray}\label{eq:kappa(0|r)}
\bar{\kappa}_i(0|r)&=&\frac{4\pi}{3} \int_{0}^r \zeta^3(x)  \frac{x^2 dx}{\sqrt{1+x^{-3}}}
\approx \frac{4\pi}{3}\zeta^3_{\infty} \int_{0}^r x^2 dx=\frac{4\pi}{9}\left(\zeta_{\infty}r\right)^3, \text{ where}\\
\zeta_{\infty}&=&\zeta(\infty)=\int_0^\infty\frac{dx}{x^2\sqrt{1+x^{-3}}}=\frac{2}{\sqrt{\pi}} \Gamma\left(\frac{1}{3}\right) \Gamma\left(\frac{7}{6}\right)
\end{eqnarray}
is the conformal time at the future infinity. Since $r$ grows exponentially with time in this era, so does the average in-degree of the oldest nodes according to Eq.~(\ref{eq:kappa(0|r)}). In the long time limit, and for degrees well below the maximum average degree, $\kappa_i\ll\bar{\kappa}_i(0|r)$, we have $r>r'\gg1$. Keeping the first two terms in the Taylor expansion of Eq.~(\ref{eq:zeta(r)-energy}) for $r\gg1$, we approximate conformal time by $\zeta(r)\approx\zeta_{\infty}-1/r$. Inserting this approximation into Eqs.~(\ref{eq:kappa(r)},\ref{eq:phi(r)}), and neglecting $x^{-3}$ there, we obtain the implicit expression for the in-degree distribution at the future infinity,
\begin{equation}\label{eq:Q(kappa,infty)}
Q(\kappa_i,\infty)=\frac{9}{4 \pi} \frac{1}{x(\kappa_i)^3 [x(\kappa_i)-1]^3},
\end{equation}
where $x(\kappa_i)$ is the solution of equation
\begin{equation}\label{eq:x(kappa)}
\kappa_i=\frac{2\pi}{9} \left[ (x-1) (2x^2-7x+11)-6 \ln{x} \right],
\end{equation}
in the region $x\ge1$. The last two equations give a nearly exact solution for the asymptotic in-degree distribution in a flat universe, because all the approximations that we have made so far become exact in the $t,r\to\infty$ limit. The only approximation that is not rigorously exact is the one due to the Laplace method in Section~\ref{sec:laplace}, yet given the astronomical value of node density $\delta$ in Eq.~(\ref{eq:planck_time}) used in this approximation, it can be also considered exact. We also note that the distribution is properly normalized since $\int_1^\infty Q(x,\infty)\left(\kappa_i\right)'_xdx=1$.

\begin{figure}
\centerline{\includegraphics[width=4in]{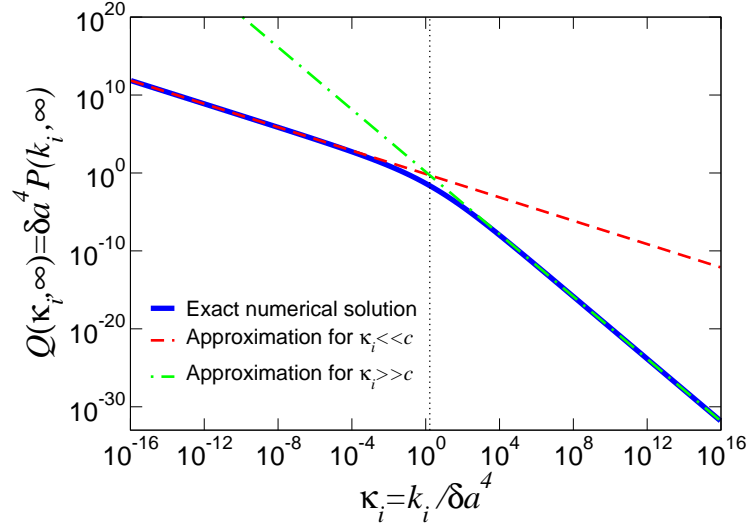}}
\caption{Exact numerical solution versus approximations for the in-degree distribution in the universe causet at the future infinity. The solid line shows the numeric solution in Eqs.~(\ref{eq:Q(kappa,infty)},\ref{eq:x(kappa)}), while the dashed(-dotted) lines are the two approximations in Eq.~(\ref{eq:P(k)approx-future}) for $\kappa_i\ll c$ and $\kappa_i\gg c$. The crossover degree value $c$ is shown by the dotted vertical line.}
\label{fig:P(k)universe_approx}
\end{figure}

Finally, we find approximations to this exact solution for small and large degrees. If $\kappa_i\ll1$, then the solution of Eq.~(\ref{eq:x(kappa)}) scales as $x(\kappa_i)=1+(3\kappa_i/\pi)^{1/4}$, whereas for $\kappa_i\gg1$, the scaling is $x(\kappa_i)=(9\kappa_i/4\pi)^{1/3}$. Substituting these scalings into Eq.~(\ref{eq:Q(kappa,infty)}), and neglecting insignificant terms there, we obtain
\begin{equation}
Q(\kappa_i,\infty)\approx\left\{
\begin{array}{ll}
\displaystyle{
\frac{3}{4}\left(\frac{3}{\pi}\right)^\frac{1}{4}\kappa_i^{-\frac{3}{4}} }
&\quad\text{if $\kappa_i \ll c$,}\\[1cm]
\displaystyle{
\frac{4\pi}{9}\kappa_i^{-2} }
&\quad\text{if $\kappa_i \gg c$,}
\end{array}
\right.
\label{eq:P(k)approx-future}
\end{equation}
where the crossover degree value $c=1.66$ is given by equating the two approximations above. Figure~\ref{fig:P(k)universe_approx} shows the exact numerical solution for the in-degree distribution at $t=\infty$, and juxtaposes it against these two approximations. The match is remarkable.

\subsection{Numeric evaluation of the analytic solution}\label{sec:numerics}

Given the results in the previous sections, the precise steps to numerically evaluate the analytic solution for the in-degree distribution in the universe at any given time are:
\begin{enumerate}
\item
For a given rescaled age of the universe $\tau$, e.g.\ $\tau=\tau_0$, compute the current value of the rescaled scale factor
\begin{equation}
r(\tau)=\sinh^\frac{2}{3}\frac{3}{2}\tau,
\end{equation}
and the normalization factor
\begin{equation}
{\cal N}(\tau)=\frac{1}{6}\left( \sinh3\tau-3\tau \right).
\end{equation}
\item
According to Eq.~(\ref{eq:zeta}), define function
\begin{equation}
\zeta(r)=2 \sqrt{r}\,\,{}_2F_1\left(\frac{1}{6},\frac{1}{2},\frac{7}{6},-r^3 \right),
\end{equation}
where $_2F_1$ is the hypergeometric function.
\item
Generate a (log-spaced) sequence of rescaled in-degrees $\kappa_i$, and for each value of $\kappa_i$ in the sequence, find numerically the solution $r^*(\kappa_i)\in[0,r(\tau)]$ of equation
\begin{equation}
\kappa_i=\frac{4\pi}{3} \int_{r^*(\kappa_i)}^{r(\tau)} \left[\zeta(x)-\zeta(r^*(\kappa_i)) \right]^3  \frac{x^2 dx}{\sqrt{1+x^{-3}}}.
\end{equation}
\item
With $r^*(\kappa_i)$ and $r(\tau)$ at hand, compute numerically the integral
\begin{equation}
\Phi(\kappa_i,r(\tau))=4 \pi \int_{r^*(\kappa_i)}^{r(\tau)} \left[\zeta(x)-\zeta(r^*(\kappa_i)) \right]^2  \frac{x^2 dx}{\sqrt{1+x^{-3}}}.
\end{equation}
\item
Finally, according to Eq.~(\ref{eq:P(k)approx}), the value of the rescaled in-degree distribution at $\kappa_i$ is given by
\begin{equation}
Q(\kappa_i,\tau)=\frac{\left[r^*(\kappa_i)\right]^4}{{\cal N}(\tau) \Phi(\kappa_i,r(\tau))}.
\end{equation}
\end{enumerate}

\subsection{Universality of $\gamma=3/4$ scaling for small degrees}\label{sec:universality}

The portion of the degree distribution with exponent $\gamma=3/4$ for small degrees below $\sim\delta a^4$ remaining in the universe causet even at long times may appear as a paradox at the first glance. Indeed, at long times, the universe is in its accelerating era dominated by dark energy, and the scale factor grows exponentially, versus polynomial growth in the matter-dominated era. Yet the degree distribution for small degrees behaves exactly the same way as in a matter-dominated universe, i.e.\ it has the same exponent $\gamma=3/4$.

The intuitive explanation of this paradox is as follows~\cite{Garriga12-private}. The degree distribution in the range of small degrees is shaped by spacetime quanta born at times $\eta'$ near the current time $\eta$. The future horizon radii of these nodes are smaller than the Hubble radius $1/H_0\sim a$. Therefore these nodes do not yet ``feel'' the acceleration of the universe. For them the universe expands as if it was matter-dominated.

To formalize this intuition we check analytically that exponent $\gamma=3/4$ for degrees $k_i\ll\delta a^4$ is universal for any scale factor. Keeping only the first term in the Taylor expansion of $\bar{k}_i(\eta'|\eta)$ at $\eta'=\eta$, we approximate
\begin{equation}
\bar{k}_i(\eta'|\eta) \approx \frac{1}{3} \pi \delta R^4(\eta) (\eta'-\eta)^4.
\end{equation}
Using this approximation to solve equation $k_i = \bar{k}_i(\eta'|\eta)$ for $\eta^*(k_i)$, and substituting the solution into Eq.~(\ref{eq:P(k)approx}), we obtain
\begin{equation}
P(k_i,\eta) \approx  \left(\frac{3}{\pi \delta} \right)^{1/4} \frac{R^3 (\eta)}{4 {\cal N}(\eta)}k_i^{-3/4},
\end{equation}
The degree distribution for small degrees is thus a power law with universal exponent $\gamma=3/4$, while the scale factor is relegated to normalization. The scale factor is important only for old spacetime quanta with large degrees, where its exponential growth in asymptotically de Sitter spacetimes is responsible for the emergence of exponent $\gamma=2$.

\pagebreak
\section{Related work}

The idea of replacing continuum spacetime with a graph or network appears in many approaches to quantum gravity, as it is a natural way to describe a discrete geometry. Causal Dynamical Triangulations (CDTs), for example, are formulated in terms of a simplicial triangulation of spacetime, which can be regarded as a (fixed valence) graph in which adjacent simplices are connected by an edge~\cite{AmJu05,AmGo08,AmJu08}. Another popular approach to quantum gravity is Loop Quantum Gravity, where the quantum states of geometry are described naturally in terms of \emph{spin networks}, which are graphs embedded into a three dimensional manifold. The edges and vertices in these graphs are colored with various mathematical structures, see~\cite{RoSp10} and references therein. There are a multitude of other approaches whose mathematical formulations have a similar discrete network-like character, such as Wolfram's evolving networks~\cite{Wolfram02}, ``quantum graphity''~\cite{KoMa08}, D'Ariano's causal networks~\cite{DaTo11}, Requardt's lumpy networks~\cite{Requardt2003}, etc.

Most descriptions of quantum spacetime geometry in terms of a graph structure are of a purely \emph{spatial} character, in that the description of space\emph{time} arises from temporal evolution of the graph whose edges usually connect spatially nearest neighbors. The intuition behind this idea is clear, being similar to the idea of approximating a continuous space by a fine lattice embedded in it. However, the precise physical meaning of the ``time'' in which the network evolves, and the manner in which the Lorentz invariant nature of special relativity can emerge from such discreteness, are often unclear~\cite{BoHe09}. Causal sets and CDTs are different in that they are described in terms of a graph-like structure which has a fundamentally space\emph{time} character.

The degree of a node is an effective measure of its age in growing complex networks. This simple observation led us here to establishing a connection between these networks and the causal networks of discrete spacetime. The notion of degree as a measure of time is reminiscent of unimodular gravity~\cite{DaLo94}.

\end{document}